\documentclass[times,5p]{elsarticle}

\usepackage[T1]{fontenc}
\usepackage{microtype}
\usepackage{lmodern}
\usepackage{booktabs}
\usepackage{amsmath,amssymb,bm}
\usepackage{xspace}
\usepackage{enumitem}
\usepackage{listings}
\usepackage{lineno}

\usepackage[pdfencoding=unicode]{hyperref}
\usepackage[subrefformat=parens,labelfont=bf,font={stretch=1.15}]{subcaption}

\usepackage[dvipsnames]{xcolor}
\definecolor{darkred}{rgb}{0.5,0,0}
\definecolor{darkblue}{rgb}{0,0,0.5}
\definecolor{firebrick}{rgb}{0.75,0.125,0.125}
\definecolor{darkgreen}{rgb}{0,0.5,0}
\hypersetup{urlcolor=darkblue,
	    citecolor=darkgreen,
	    linkcolor=firebrick}

\usepackage{cleveref}
\usepackage[version=4]{mhchem}
\usepackage[object=vectorian]{pgfornament}
\usepackage{siunitx}[=v2]

\journal{Astroparticle Physics}

\newcommand{\xmax}{\ensuremath{X_{\rm max}}\xspace}
\newcommand{\gcm}{\ensuremath{{\rm g/cm}^2}\xspace}
\newcommand{\ecr}{\ensuremath{E_{\rm CR}}\xspace}
\newcommand{\edeposit}{\ensuremath{E_{\rm deposit}}\xspace}
\newcommand{\ecore}{\ensuremath{E_{\rm core}}\xspace}
\newcommand{\eapparent}{\ensuremath{E_{\rm apparent}}\xspace}

\begin{document}
\begin{frontmatter}

\title{In-ice Askaryan Emission from Air Showers: \\ Implications for Radio Neutrino Detectors}
\cortext[cor1]{Email address: christian.glaser@physics.uu.se}

\author[a]{Alan Coleman}
\author[a]{Christian Glaser}
\author[b]{Ryan Rice-Smith}
\author[b]{Steven Barwick}
\author[c]{Dave Besson}

\affiliation[a]{Uppsala University Department of Physics and Astronomy, Uppsala SE-752 37, Sweden.}
\affiliation[b]{Dept. of Physics and Astronomy, University of California, Irvine, United States}
\affiliation[c]{University of Kansas, Dept. of Physics and Astronomy, Lawrence, KS 66045, USA}

\begin{abstract}
One of the most promising techniques for detecting ultra-high energy neutrinos involves the use of radio antennas to observe the 10--1000\,MHz radiation generated by the showers that neutrinos induce in large volumes of ice. The expected neutrino detection rates of one neutrino or less per detector station per 10 years make the characterization of backgrounds a priority. 
The largest natural background comes from ultra-high energy cosmic rays which are orders of magnitude more abundant than neutrinos. Particularly crucial is the understanding of geometries in which substantial energy of the cosmic-ray-induced air shower is deposited in the ice giving rise to a compact in-ice shower close to the ice surface. We calculated the radio emission of air-shower cores using the novel CORSIKA~8 code and found it to be similar to the predictions for neutrino-induced showers. For the first time, we calculated the detection rates for $\mathcal{O}$(\SI{100}{m}) deep antennas yielding 10-100 detections per year and detector station, which makes this a useful calibration source as these downward-going signals can be differentiated from neutrino-induced showers based on the signal arrival direction. 
However, the presence of reflection layers in the ice confuses the arrival directions, which makes this a potentially important background. We review the existing information on reflecting layers in the South Pole glacier and, for the first time, quantify the corresponding rate of reflected air-shower signals for the proposed IceCube-Gen2 radio array and discuss mitigation strategies. The reflectivity of the layers is the dominant uncertainty resulting in rate predictions of much less than one detection to several detections per year for IceCube-Gen2 if not mitigated. 
\end{abstract}
\end{frontmatter}


\section{Introduction}
Detection of neutrinos at ultra-high energies (UHE, $E >\SI{e17}{eV}$) would open a new window to the most violent phenomena in our universe \cite{Ackermann:2019cxh,Ackermann:2019ows,Ackermann:2022rqc}.
The detection of neutrinos is challenging, though, due to their small interaction cross-section and the rapidly decreasing neutrino flux towards higher energies. Radio detection remains the most promising technique at these energies, allowing for cost-efficient instrumentation of large volumes~\cite{Barwick:2022vqt}. In particular, the in-ice radio detection technique, which instruments ice sheets with arrays of compact radio detector stations, is at an advanced development stage. The technology has been successfully developed and explored with the pilot arrays RICE, ARA and ARIANNA at the South Pole and Moore's Bay in Antarctica~\cite{Kravchenko:2011im, ARA2011, ARIANNA:2019scz}. The ongoing construction of the Radio Neutrino Observatory in Greenland (RNO-G) makes the detection of the first UHE neutrino a possibility in the coming years~\cite{RNOGWhitePaper2021}, and the planned radio array of IceCube-Gen2 will increase the sensitivity by another factor of 10~\cite{IceCube-Gen2-TDR}. 

The flux of UHE neutrinos originates from interactions of the well-established flux of UHE cosmic rays (CRs) with the cosmic microwave background, as well as other production mechanisms (see \cite{Valera:2022wmu} for an extensive overview of theoretical models). However, while this flux is guaranteed, the uncertainties in the composition and overall normalization of the UHE cosmic-ray flux result in an unclear picture of the corresponding flux of UHE neutrinos. Combined with other potential mechanisms of neutrino production at these energies, the total flux on Earth is even less constrained.

Consequently, a decisive observation of this flux will require a precise understanding of all relevant backgrounds. The physical background at UHE is dominated by secondary particles generated by cosmic-ray interactions in the atmosphere. The impulsive radio-frequency signals generated by these secondaries can mimic those of UHE neutrino interactions in the ice through various channels, and efforts to reduce these backgrounds require a detailed understanding of the signature of each mechanism. While radio emission from two of these channels, in-air radio emission which originates mainly from geomagnetic emission from opposing deflection of $e^\pm$ during the air-shower development \cite{KahnLerche1966, Scholten200894,Huege2016} and stochastic energy deposits in the ice by $>$\,$10^{16.5}$\,eV muons generating Askaryan emission have been well studied~\cite{Garcia2020,Pyras:2023crm}, the third channel has only recently been explored~\cite{DeKockere:2022bto,ryan_rice_smith_2022_6785120,DeKockere:2024qmc}: the in-ice Askaryan emission from the air-shower cores that reach the air-ice boundary \cite{JavaidPhD2012,deVries:2015oda}. In the following, we focus on this background process. We refer the reader to the review of \cite{Barwick:2022vqt} for an overview and a more detailed description of the other relevant background processes for in-ice radio neutrino detection. 

Askaryan emission from air-shower cores has additional modes of producing background that may be more difficult to reject and would require a better understanding of the ice properties and more generally concentrated calibration efforts compared to the other background channels. Reflection layers in the ice, inhomogeneities of the bulk ice-crystal structure, can reflect signals back up that appear to originate from deep in the ice, as expected from UHE neutrino interactions. Paired with the abundant flux of cosmic rays, the background rate can be substantial. Here, we quantify the expected background rate for the first time and discuss its mitigation strategies.  

In-ice radio detector stations are built by either installing LPDA antennas in slots close to the surface of the ice, referred to as \emph{shallow} in the following, or by drilling narrow holes down to 100-200~m depth, which are instrumented with dipole and slot antennas that fit the borehole, referred to as \emph{deep}. 
The main focus of our predictions is the envisioned radio array of IceCube-Gen2~\cite{IceCube-Gen2-TDR, IceCube-Gen2:2021rkf}, which comprises a hybrid array of \emph{shallow} and \emph{deep} station components spread over an area of \SI{500}{km^2}. The shallow part is triggered by four downward facing LPDA antennas by requiring time-coincident high/low threshold trigger in two out of the four antennas. The deep component is triggered by an interferometric phased array consisting of four dipole antennas placed above each other at minimal separation~\cite{ARA2019-PA}. The dominant noise source is irreducible thermal noise fluctuations, and the trigger threshold is adjusted to generate a \SI{100}{Hz} trigger rate. The trigger is calculated in a lower bandwidth of approx.~80-\SI{220}{MHz} than the final readout, as this has been shown to increase the sensitivity to neutrinos \cite{Glaser2020Bandwidth}.

\begin{figure*}[tb]
    \centering
    \includegraphics[width=\textwidth]{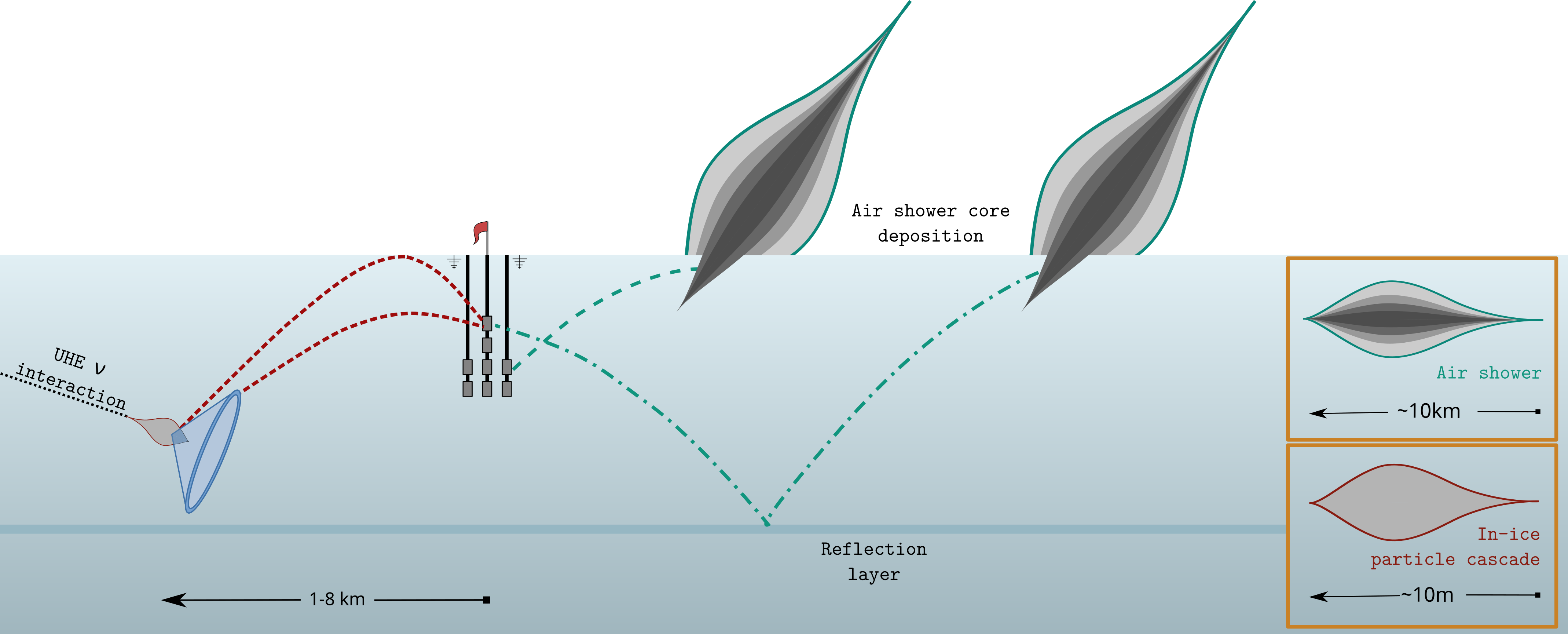}
    \caption{The radio-frequency emission from the cascades, that are generated by neutrino interactions, follows a curved path through the ice (dotted red line) and can be seen by antennas up to kilometers away. One of the major backgrounds for identifying neutrinos comes from air-shower cores impacting the ice. While this emission would normally immediately propagate to the antennas from the surface (dashed green line), in the presence of a reflecting layer, the arrival direction would appear to come from the deep ice, instead (dash-dotted green line).
    }
    \label{fig:background_diagram}
\end{figure*}

An illustration of the background is provided in Fig.~\ref{fig:background_diagram}. 
Cosmic rays induce huge particle cascades in the atmosphere, typically referred to as air showers. The shower might not be fully developed when reaching ground level, so the particle cascade continues to develop in the ground. Especially at high-altitude sites, such as the typical locations for in-ice radio neutrino detectors at the South Pole and central Greenland, and vertical air showers, a large fraction of the initial cosmic-ray energy is deposited in the ground. Furthermore, a substantial part of the air-shower energy is concentrated close to the shower axis \cite{DeKockere:2022bto}, thus giving rise to a compact in-ice shower. As we will show in Sec.~\ref{sec:C8} using the novel CORSIKA 8 code \cite{CORSIKA:2023jyz, Alameddine:2024cyd}, this in-ice shower will emit Askaryan radiation, similar to the in-ice showers expected from neutrino interactions in the ice. 

First, this offers a calibration signal for in-ice radio detector signals, which will allow probing of the trigger efficiency, signal identification, and reconstruction algorithms under realistic conditions. The radio emission will reach the deeper antennas of an in-ice radio detector station on a direct path. This geometry is of limited concern as a background because the signal arrival directions are restricted to coming from above with zenith angles in a given range with small overlap to the expected signal arrival directions from neutrino-induced in-ice showers. We will later calculate that the expected detection rates are large, with tens to hundreds of detections per station per year. Once the air-shower core Askaryan emission is understood and modeled accurately, it will be a useful calibration signal to demonstrate that a \emph{deep} detector system is working. The full analysis pipeline responsible for triggering on neutrinos and distinguishing genuine signals from the overwhelming background noise can be systematically validated. Even the reconstruction of the energy and shower direction can be probed on a statistical basis, as we can predict the expected energy and arrival direction distributions with high accuracy. In addition, for a subset of the events, the in-air radio emission generated by the air shower can be detected by the upward-facing LPDA antennas of the \emph{shallow} detector component, which will provide an independent measurement of the air-shower direction and an estimate of the energy deposited in the atmosphere \cite{PierreAuger:2015hbf,PierreAuger:2016vya,Welling:2019scz}.

Second, natural ice sheets contain reflection layers (see next section) that act as a mirror and reflect the signal back up so that it appears to originate from below. The reflectivity is small and at most 1\%, but because cosmic rays are much more abundant than neutrinos, have a much larger interaction cross-section, and the cosmic-ray energy spectrum extends to high energies, the detection rate of \emph{reflected shower cores} might still be substantial. In this article, we quantify the background detection rate, characterize the background, and discuss mitigation strategies.

Although measurements indicate that the ice at Summit Station Greenland (the RNO-G site) does not have reflection layers with sufficiently high reflectivity to cause significant background rates~\cite{RNO-G:2023Ice}, RNO-G can probe the Askaryan emission of air-shower cores through these direct signal trajectories from the shower core to its \SI{100}{m} deep antennas. Hence, we also make predictions for RNO-G. Similar to Gen2, RNO-G is triggered by a phased array with four dipole antennas installed at a depth of \SI{100}{m} \cite{RNOGWhitePaper2021}. The trigger thresholds are adjusted to a noise trigger rate of \SI{1}{Hz} as this is the maximum data bandwidth the DAQ can handle. We note that also for Gen2, the maximum data bandwidth to send and store events is \SI{1}{Hz}, however, through a second trigger stage on station level, the data rate can be reduced to \SI{1}{Hz} without a significant loss of signal efficiency \cite{Arianna:2021vcx,RNO-G:2023oxb}.

The paper is structured as follows: In section two, we review the current knowledge of reflection layers at the South Pole. In section three, we calculate the radio emission from in-ice shower cores. In section four, we calculate the detection rates before we discuss mitigation strategies in section five and discuss our findings in section six.

\section{Review of Reflection Layers}
\label{sec:ice}
Typically, reflective layers within the ice sheet are attributed to originate in one of four sources. First, yearly summer snow melt forms crusts with a slight overdensity relative to surrounding snow, which then increasingly embed within the ice sheet. The density contrast is typically not more than 1\% for snow melt layers, which then embed within, and below the firn. Such yearly crusts are unlikely to produce significant reflections. Second, episodic processes such as sulfate depositions resulting from terrestrial volcanism can lead to thin (thicknesses typically less than one skin depth) but highly conductive layers. These are often visible during inspection of ice cores, with correspondingly high reflectivity and are perhaps most likely responsible for the internal echoes observed in radar surveys. Third, re-alignment of the crystal orientation fabric, over the course of many years, can also result in an evolution of the dielectric permittivity over macroscopic distances within the ice sheet. Finally, depending on wind patterns, sea salts can be blown inland, resulting in fluctuations in the local $Na^+$ or $Cl^-$ ionic (and therefore conductive) concentrations. 

Information on internal layers at the South Pole may be derived from multiple sources. Dedicated aerial radar surveys of the polar icecaps by the British Antarctic Survey (BAS) and the Center for Remote Sensing and Information Systems (CReSIS)~\cite{li2012high} not only provide large-scale ice sheet thickness data, but also allow estimates of internal layer reflectivity, albeit typically over fairly narrow bandwidths; synthetic aperture radar (SAR) techniques are then used in post-processing to accentuate the visibility of both the bed reflection, and any internal layering. Impulsive broadband signals (with time duration of $\cal{O}$(10~ns)) are complementary to narrowband aerial surveys and provide extended frequency-domain information with high spatial resolution, but are limited to surveys over a small surface area.

A recent study~\cite{besson2023polarization} re-calculated internal layer reflectivity measurements drawn from archival RICE experimental data, summarized in Tab.~\ref{table:ReflectivityLayers}. Interestingly, reflection coefficients were obtained that were approximately 5\,dB stronger for signal polarizations parallel to the local ice flow direction, although the possibility that this may be an ice effect (birefringence, e.g.) has not yet been ruled out. To reduce antenna-dependent systematics, the layer reflectivity calculations are normalized to the bedrock reflection. A bedrock reflectivity of $R=1$ is expected if the bottom surface is smooth,  and the temperature of the ice is above the melting temperature (warmer than -2$^\circ$~C) such that there is a layer of (impurity-free) water below the ice sheet. The IceCube collaboration has measured the temperature profile of South Polar ice and estimated that the temperature at bedrock is -3.43$^\circ$~C~\cite{icecubeTemp}, close to the melting point. A more recent calculation of the ice temperature profile at the South Pole concludes that "the ice-bed interface is thawed" and includes a revised value for the ice thickness of 2,880~m\cite{hills2022geophysics}. Furthermore, this paper includes a survey of basal reflected power in the vicinity of the South Pole, and shows that the average reflected power differs by only 1~dB relative to a nearby subglacial lake, which has reflectivity very close to 1.   Assuming unity for the bed reflectivity, the maximal reflectivity (maximum allowed by the errors) of all internal layers listed in Tab.~\ref{table:ReflectivityLayers} is approximately -40~dB. We note that these measurements are typically made at normal incidence; since the reflecting layers are generally less than one skin depth, oblique incidence would be expected to yield a higher reflectivity simply owing to a larger layer chord traversed.

\begin{table} 
\centering

\begin{tabular}{c c c}
\hline \hline
$t_{layer}$ & $d_{layer}$ & $R_{layer}/R_{bed}$ (dB)\\ \hline 
6009 ns &  $521\pm11.5$ m & -48+/-9 \\
9674 ns &  $829\pm17.4$ m & -47+/-8 \\
9792 ns &  $840\pm18$ m & -46+/-8 \\
9905 ns &  $849\pm18$ m & -54+/-12 \\ \hline \hline
\end{tabular}
\caption{Estimated depth and reflectivity of internal layers measured at the South Pole from  ~\cite{besson2023polarization}. The first column indicates the echo time of the measured layer, followed by the estimated depth of the layer and, finally, the reflectivity of the layer relative to the bedrock.}
\label{table:ReflectivityLayers}
\end{table}

In another study, bi-static vertically-penetrating radio pulses at \SI{380}{MHz} were transmitted into South Polar ice as tone bursts of \SI{400}{ns} duration ~\cite{Barwick:JGR2005}. The attenuation length of the ice was determined from time-integrated signals that were reflected from the ice-rock interface at the bottom of the ice sheet. However, the reflected power was collected as a function of depth throughout the ice sheet. These data show prominent peaks at depths of \SI{466}{m} and \SI{800}{m}, corroborating the existence of internal reflection layers. We use the time-integrated reflected power as a function of depth in the ice from that study to calculate the reflectivity of a given layer using an ice attenuation profile appropriate to the South Pole and the assumption that the reflection coefficient of the reflecting surface at the bottom of the ice sheet has $R=1$.   We use the measured ratio of reflected power at a given depth to the reflected power from the bottom of the ice sheet, and the expected attenuation in power during propagation, to calculate $R_{layer}/R_{bed}$, which for both depths is $-45\,{\rm dB}\pm5\,{\rm dB}$, consistent with Table~\ref{table:ReflectivityLayers}. We note that the depth estimate between different measurements can have significant systematic uncertainties due to the choice of the density/index-of-refraction profile that is used to convert the measured time delays to depth. 

In both studies cited in the two preceding paragraphs, it was not possible to identify internal reflection layers at depths shallower than $\sim$400~m because pulsed radio power from the transmitter was scattered by the firn ice near the surface to the receiver antenna that consequently saturated the amplifiers.  The minimum depth of observable reflection layers was correlated with the decay time of the amplifier back to normal linear response. However, it is possible that one or more reflection layers can exist at shallower depths. 

In summary, there are clear indications of reflection layers at the South Pole, however, current measurements of their properties, in particular their reflectivity, have large uncertainties. New measurement campaigns or reanalysis of existing data sets are required to better constrain the properties of reflection layers at the South Pole. For now, we will simulate the effect of reflection layers at \SI{300}{m}, \SI{500}{m} and \SI{830}{m} depth on detection rates of reflected Askaryan signals from air shower cores to assess the importance of the reflector depth. Throughout our study, we keep the reflectivity as a free parameter to determine how the background rate scales with reflectivity. This will allow us to estimate the importance of reflection layers on in-ice neutrino detection, which, in turn, will inform how important further studies of reflection layers are (see our discussion in Sec~\ref{sec:discussion}). 

\section{Calculation of in-ice radio emission from shower cores}
\label{sec:radio_emission_sims}

The characteristics of the Askaryan emission generated from in-ice air-shower cores depends on the energy content still remaining in the air shower and the amount of lateral spread of these particles on the ground. While the former results in an overall normalization of the generated electric field, the latter defines the frequency content. In this section, we describe a first-order method for estimating the amount of in-ice radio emission from air shower cores. 

The particle content of air showers is dominated by the electromagnetic component, which contains $>$\,95\% of the energy after a few electromagnetic interaction lengths. The growth and attenuation of this component, i.e., the longitudinal shower profile, has been directly studied at ultra-high energies using the fluorescence imaging technique by the Pierre Auger (Auger), Telescope Array (TA), and Yakutsk observatories~\cite{PierreAuger:2018gfc,PierreAuger:2023bfx,PierreAuger:2023kjt,Abbasi:2014sfa,TelescopeArray:2018xyi,TelescopeArray:2020bfv,Knurenko:2019oil}. Due to its larger instrumented area, the increased statistics at Auger allow for the highest statistical precision of measuring the energy evolution of both the shape of the longitudinal profile and the penetration depth. In this work, we use the description of the energy evolution of the depth of shower maximum (\xmax) from the most recent Auger results~\cite{PierreAuger:2023bfx}. Similarly, the shape of the longitudinal profile was measured and parameterized~\cite{PierreAuger:2018gfc}. Together, these fully describe the shower profile as a function of cosmic-ray energy:
\begin{multline}
    A(X, \xmax, E) = \left(1 + R(E) \frac{X - \xmax}{L(E)}\right)^{R(E)^{-2}} \\ \times  \exp \left( - \frac{X - \xmax}{R(E)\,L(E)} \right) \, ,
    \label{eq:gaisser_hillas_LR}
\end{multline}
where $X$ is the atmospheric depth (integrated atmospheric density), $E$ is the cosmic-ray energy, and $L$ and $R$ are parameterizations derived from data~\cite{PierreAuger:2018gfc}. By combining the measurements of the evolution of \xmax with energy~\cite{PierreAuger:2023bfx}, the average shower profile can be determined.
From this parameterization, we can estimate the energy deposited into the ice by integrating~\cref{eq:gaisser_hillas_LR} from the slant depth at the ground ($X_{\rm ground}$) to infinity. $X_{\rm ground}$ depends on the zenith angle, $\theta$, and scales, to first order, as $X_{\rm ground} = X_{\rm ground, vertical} / \cos \theta$. Combined with the energy of the primary cosmic ray, the energy that arrives at the ground is given by,
\begin{equation}
    \edeposit = \frac{E}{N(\xmax, E)} \int^{\infty}_{X_{\rm ground}}A(X, X_{\rm max}, E)\, dX, \label{eq:deposited}
\end{equation}
where $N(\xmax, E)$ is a normalization constant given by the integral of~\cref{eq:gaisser_hillas_LR} over all $X$.
\begin{figure}[tb]
    \centering
    \includegraphics[width=0.99\columnwidth]{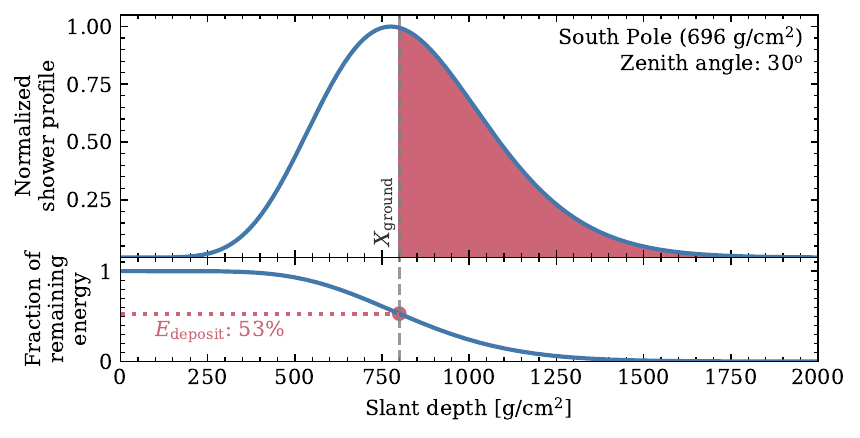}
    
    \caption{
        For the longitudinal profile shown in the upper panel, the amount of deposited energy on the ground is taken as the remaining air shower content beyond $X_{\rm ground}$. The inverse CDF is shown in the bottom panel. For such a shower, with an \xmax of 775\,\gcm, which is typical for cosmic rays with energies of $\sim$\,$10^{19}$\,eV, the remaining shower content at the ground at the South Pole is 53\%.
        \label{fig:longitudinal_profile}
    }
\end{figure}
This function is shown schematically in the lower panel of Fig.~\ref{fig:longitudinal_profile}.

The remaining energy that penetrates a few tens of meters into the ice will produce a radio signal that can be observed by in-ice radio antennas. However, the lateral spread of the electromagnetic particles upon reaching the air-ice interface will result in a loss of coherence with respect to a shower that is initiated at this location with energy \edeposit by a single particle, e.g., a neutrino interaction. As we will later use Askaryan emission models that were developed for neutrino-induced in-ice showers, we need to correct for this loss of coherence. The observed air-shower core signal will be smaller by a factor $f$, giving an effective energy of the air-shower core, $\ecore = f \times \edeposit$, as the amplitude of the Askaryan signal scales linearly with the shower energy. The work of \cite{DeKockere:2022bto} found significant radio emission from in-ice shower cores similar to the emission expected from neutrino-induced in-ice showers. In that work, air shower simulations with CORSIKA~7 were combined with in-ice shower simulations using GEANT-4 to simulate the in-ice radio emission, which was named FAERIE simulations by the authors. Although no frequency spectrum was shown, the time-domain pulses are very short $\mathcal{O}$(\SI{1}{ns}), indicating high-frequency content $\mathcal{O}$(\SI{1}{GHz}). Furthermore, we note that radio neutrino detectors optimize the trigger bandwidth to low frequencies of approx.~80-\SI{200}{MHz} compared to the full detector bandwidth of up to \SI{1}{GHz}, which has been shown to increase the sensitivity to neutrinos \cite{Glaser2020Bandwidth}. Hence, a loss in coherence of shower cores at high frequencies is less relevant when it comes to triggering a radio neutrino detector, and for calculating detection rates, our simplified signal modeling only needs to match the true signals in the 80-\SI{200}{MHz} frequency band. 

\subsection{CORSIKA 8 simulation of the radio emission from air-shower cores}
\label{sec:C8}
The recent development of the CORSIKA 8 code \cite{CORSIKA:2023jyz,Alameddine:2024cyd,Engel:2018akg} allows us to conveniently simulate the in-ice radio emission from air-shower cores in a single framework (in contrast to FAERIE). We simulated several vertical air showers with proton primaries of various energies, ranging from \SI{e15}{eV} to \SI{e19}{eV}. We approximate the ice by a homogeneous medium with an index-of-refraction of 1.78, without attenuation. The observer positions are placed at a depth of \SI{100}{m} and with horizontal displacements from the shower axis to map relevant viewing angles around the approximate Cherenkov cone with a \SI{1}{\degree} spacing. We only calculate the radio emission from in-ice shower tracks, i.e., we don't show and consider the in-air radio emission in this work. 
\begin{figure*}[t]
    \centering
    \includegraphics[width=\textwidth]{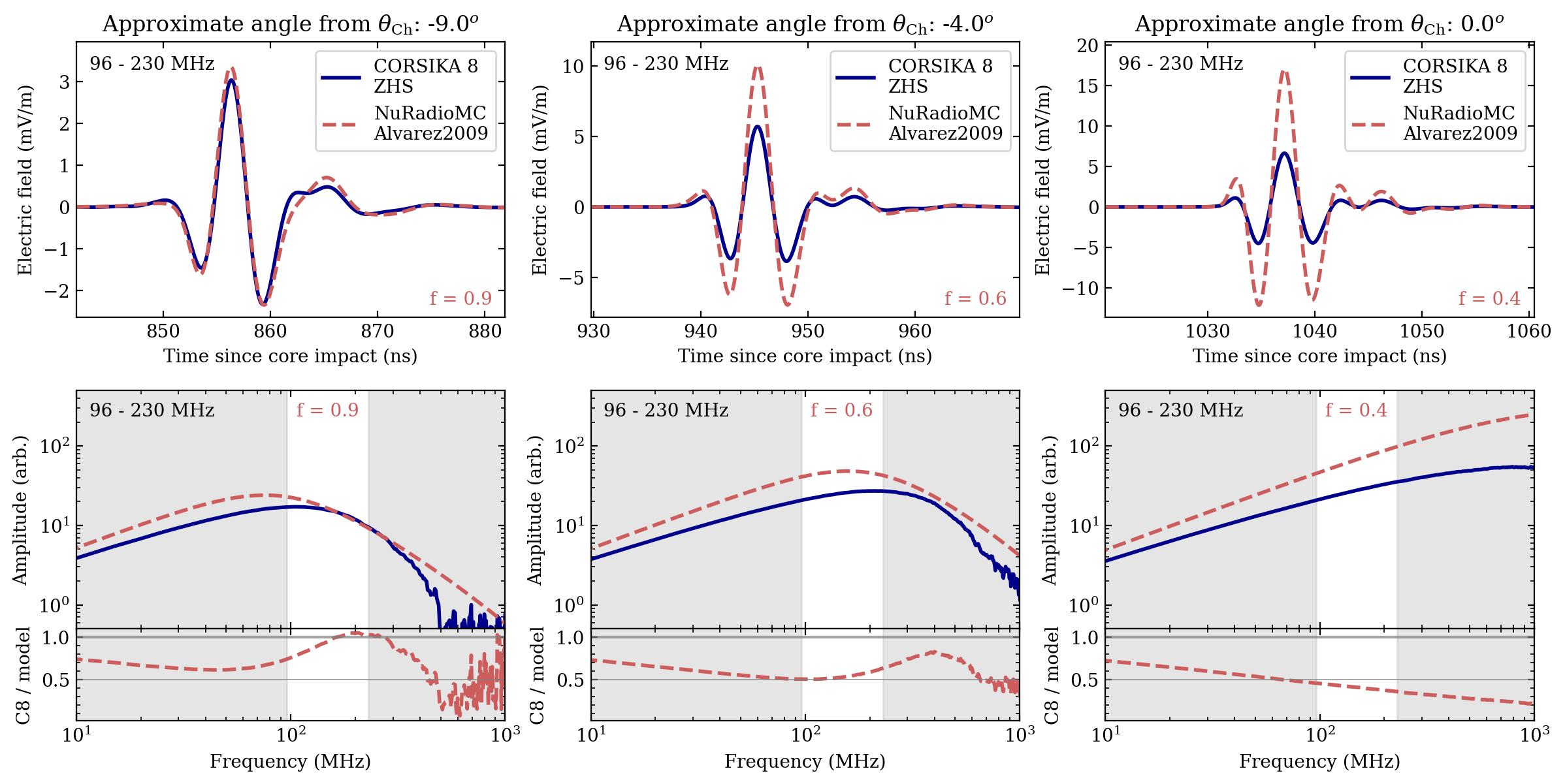}
    \caption{Radio emission from air-shower cores simulated by CORSIKA 8 using the ZHS algorithm. A vertical air shower with a \SI{1}{EeV} proton primary was simulated. The upper panels show the radio signal observed at a depth of \SI{100}{m} at three different viewing angles, filtered to the trigger bandwidth of 96-\SI{230}{MHz}. Times are given w.r.t. when the core impacts the air-ice interface. The lower panels show the corresponding frequency spectra, with the bandpass region highlighted in white. Also shown is the prediction of the radio emission of a neutrino-induced in-ice shower with the same deposited energy and approximately the same viewing angle using the Alvarez2009 model \cite{Alvarez-Muniz:2009jsq, Alvarez-Muniz:2010hbb,Glaser:2019cws} The lower panels show the ratio between the CORSIKA 8 prediction and this neutrino-Askaryan emission model.}
    \label{fig:C8-1EeV}
\end{figure*}

In Fig.~\ref{fig:C8-1EeV}, we show the in-ice radio emission generated by a vertical air shower initiated by a \SI{1}{EeV} proton at the South Pole. In this example, \edeposit = $10^{17.8}$~eV is deposited into the ice. The time-domain signals are filtered to the trigger bandwidth and shown for three different viewing angles. The right panel shows the observer with the largest signal strength, i.e., the observer which is closest to the  Cherenkov cone. The center and left panels show the emission at two selected directions at \SI{4}{\degree} and \SI{9}{\degree} away from the observer with the largest signal strength. 

We note that our approach, which considers only the in-ice portion of the particle tracks, in principle introduces additional contributions from the sudden appearance of charges--a contribution to transition radiation (see e.g. \cite{Motloch:2015wca}). However, our results do not show any indications of such effects. Instead, the observed emission is consistent with expectations from Cherenkov radiation: the amplitude increases with frequency up to a cut-off frequency that peaks at the Cherenkov angle and rapidly decreases away from it. This supports the validity of our current approach. Nonetheless, it would be interesting to investigate potential contributions from transition radiation in future studies.

In Fig.~\ref{fig:C8-1EeV}, we also present predictions for the radio emission from a neutrino-induced in-ice NC shower with a shower energy of \edeposit, using the parameterization from \cite{Alvarez-Muniz:2009jsq, Alvarez-Muniz:2010hbb}, commonly referred to as \emph{Alvarez2009} and available through NuRadioMC \cite{Glaser:2019cws}. The comparison, however, is complicated by the fact that the Cherenkov angle is not well-defined for nearby showers. In contrast, neutrino-induced in-ice showers typically occur at distances on the order of $\mathcal{O}$(\SI{1}{km}), where they can be reasonably approximated as point-like emitters \cite{Glaser:2019cws}. To achieve the most accurate comparison, the \emph{Alvarez2009} parameterization is evaluated at the Cherenkov angle for the rightmost panel of Fig.~\ref{fig:C8-1EeV}, and at \SI{-4}{\degree} and \SI{-9}{\degree} from the Cherenkov angle for the other two panels. Since our CORSIKA 8 simulations placed observers at \SI{1}{\degree} intervals along the viewing angle, the observer with the maximum signal strength can be at most \SI{0.5}{\degree} away from the Cherenkov cone.

Within the narrow trigger band, we find that the time-domain waveforms are very similar in shape. As expected, the amplitudes from the shower cores are reduced compared to the neutrino-induced in-ice showers, with an $f$-factor of 0.4 near the Cherenkov cone. This factor increases to $f = 0.6$ at \SI{4}{\degree} off-cone and reaches $f = 0.9$ at \SI{9}{\degree} off-cone. We also observe that the cutoff frequency at the approximate Cherenkov angle for shower cores occurs at lower frequencies than for in-ice showers. However, the decline in cutoff frequency with increasing off-cone angle is slower for shower cores, leading to an inversion at larger off-cone angles. Consequently, the Cherenkov cone is broader, and the $f$-factor increases with the viewing angle.

By analyzing air showers of different energies, we find that the $f$-factor is most closely tied to the shower age at the surface, which we quantify by the ratio \edeposit/\ecr which captures both the zenith angle and \xmax dependency. We observe that the f-factor is generally larger for viewing angles further from the Cherenkov cone. For a vertical \SI{e17}{eV} proton-induced air shower, we find $f$-factors between 0.3 to 0.7, and for a 20$^\circ$ \SI{e19}{eV} air shower, we find $f$-factors between 0.2 and 0.7. While we take into account that less energy is deposited into the ice for inclined air showers through Eq.~\ref{eq:deposited}, we expect the $f$-factor to decrease with larger zenith angles because the shower reaches the ice at a later shower age with increased lateral spread.

While more work needs to be done to properly parameterize the radio emission from air-shower cores, the results from the FAERIE simulations \cite{DeKockere:2022bto,DeKockere:2024qmc} and the new CORSIKA 8 results presented here, indicate that average $f$-factors of around 0.1--0.9 are plausible. 
In future work, we plan to incorporate the effects of an inhomogeneous ice sheet where the index-of-refraction transitions from approx.~1.4 at the surface to 1.78 at deeper depth into CORSIKA 8 to make more accurate predictions. Then, the emission can be parameterized similar to what has been done for neutrino-induced showers, e.g., \cite{Alvarez-Muniz:2005mez}, which can then directly be used in NuRadioMC to simulate the effective area to air-shower cores. We note that using CORSIKA 8 simulations directly to calculate the effective area is not feasible because of the long runtimes. A single CORSIKA 8 shower simulation with the calculation of radio emission can take up to a month, but the calculation of the effective area with sufficient statistical precision requires the simulation of hundreds of thousands of showers, particularly at low energies where the trigger probability is small (see Sec.~\ref{sec:effective_area}).

In the following, we present all results as a function of the $f$-factor, enabling detection rates to be determined for any given $f$-value. We recognize that using a single scaling factor, $f$, is a simplification, as the degree of decoherence depends on multiple parameters, including shower age at the surface, energy, \xmax, and zenith angle. Nevertheless, an effective $f$-factor can be defined by averaging over the relevant phase space. When a specific $f$-value is required for presenting results, we adopt $f = 0.3$, which is our best estimate for the effective value. The key finding from the study above is that the uncertainty introduced by the $f$-factor approximation leads to variations by a factor of 2–3, but not by orders of magnitude.

\section{Calculation of air-shower core detection rates}

\begin{figure}[tb]
    \centering
    \includegraphics[width=0.99\columnwidth]{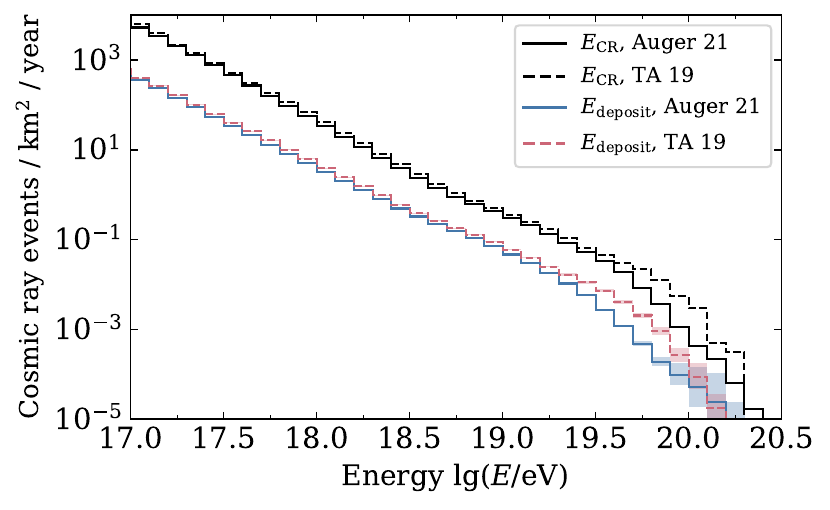}
    \caption{
        The flux of the energy of cosmic rays (black lines) and the amount of energy that they deposit on the ground at the South Pole (colored lines) are shown, assuming the flux measured by Auger (solid) and TA (dashed). The bands show the 68\% confidence interval, which combines the uncertainties from the measured cosmic-ray flux and \xmax values.
    \label{fig:deposited_energy}
    }
\end{figure}

The detection rate of air-shower cores is calculated using the following procedure. We first provide a summary and then describe the steps in detail.
\begin{enumerate}
    \item The measured cosmic-ray energy spectrum and \xmax distributions are used to calculate the energy spectrum of deposited energy according to Eq.~\eqref{eq:deposited} (Fig.~\ref{fig:deposited_energy} and \ref{fig:deposited_energy_cosine}).
    \item The effective area of an in-ice radio detector station to shower cores for reflection layers at various depths is simulated using NuRadioMC \cite{Glaser:2019cws}. This yields a parameterization of the effective area as a function of shower energy and incident zenith angle for each reflection layer, as well as for the direct signal path to the receivers (see Fig.~\ref{fig:effective_area_direct} and \ref{fig:effective_area}).
    \item The deposited energy spectrum (step 1) is folded with the effective area (step 2) to obtain the detection rate of shower cores (Fig.~\ref{fig:reflected_events_vs_Ecr} and following). 
\end{enumerate}

\subsection{Calculation of deposited energy spectrum}

Above $10^{19}$\,eV, the measurement of the flux of cosmic rays by the Pierre Auger Observatory and the Telescope Array includes some tension, resulting from a systematic difference in the energy assignment of air showers using the fluorescence techniques~\cite{PierreAuger:2023wti}. Therefore, we perform the calculation twice, once for the energy spectrum measured by Auger~\cite{PierreAuger:2021hun}, and once for the energy spectrum measured by TA~\cite{Ivanov:2020rqn}. 
For the distribution of \xmax we solely use the measurement of the Pierre Auger Observatory due to its better precision and unbiased presentation, i.e., all detector effects were unfolded. We use the measurement of the first two moments of the \xmax distribution, i.e., mean and standard deviation from~\cite{PierreAuger:2023kjt}. 

\begin{figure}[tb]
    \centering
    \includegraphics[width=0.99\columnwidth]{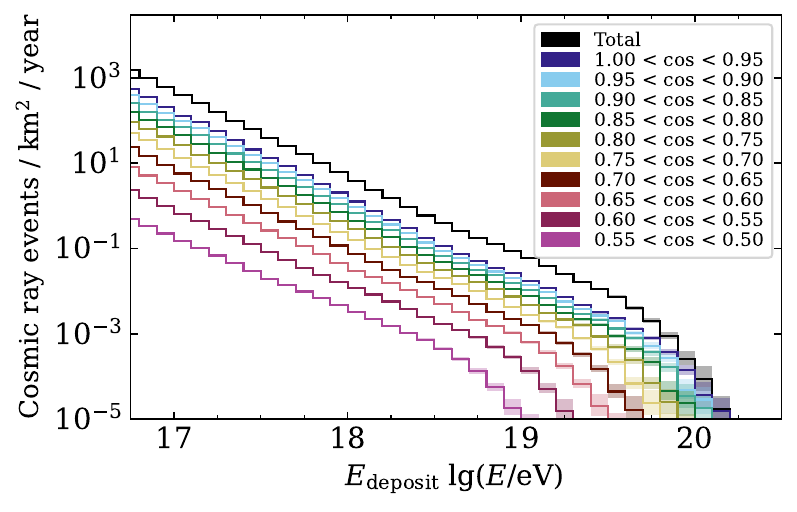}
    \caption{
        The flux of shower cores is shown as a function of the remaining energy at ground level for several zenith-angle bands. The flux of cosmic rays is assumed to be the one measured by TA.}
    \label{fig:deposited_energy_cosine}
\end{figure}

We perform the analysis in small bins of cosmic-ray energy of 0.1 in $\log_{10}(E)$ and cosine zenith angle of 0.05. For each cosmic-ray energy bin, we draw 200 \xmax realizations from a normal distribution with mean and sigma according to the measured values and calculate the deposited energy using Eq.~\ref{eq:deposited}. Additionally, the statistical measurement uncertainties of these quantities are incorporated by randomly shifting the central values of the energy spectrum and \xmax distributions within the reported errors. Where necessary, the measured values are extrapolated to cover the relevant phase space. This yields a cosmic-ray energy spectrum as a function of deposited energy, as shown in Fig.~\ref{fig:deposited_energy}. At South Pole altitude, the deposited-energy spectrum is offset from that of cosmic rays by about 0.5 in $\log_{10}(E)$.  Fig.~\ref{fig:deposited_energy_cosine} shows the deposited-energy spectrum for different cosine zenith bands. Each band covers the same solid angle in the sky. As expected, the flux is highest for vertical showers and decreases for more inclined showers as \xmax is further away from the ground and less energy reaches the ground (the atmospheric depth approximately increases with $1/\cos(\theta)$). Hence, vertical showers contribute the most to the flux, for a fixed value of \edeposit, by over three orders of magnitude compared to one at $60^\circ$. 

We propagate the statistical uncertainties of the spectrum and \xmax measurement throughout the entire analysis and show it as bands around the nominal values in all plots. 

\subsection{Calculation of effective area to shower cores}
\label{sec:effective_area}
With the NuRadioMC simulation code \cite{Glaser:2019cws}, we can simulate the effective area to in-ice showers, in particular, the showers initiated by the air shower cores beneath the ice surface. We simulate the exact detector as foreseen for IceCube-Gen2 with the same simulation settings as in previous studies \cite{IceCube-Gen2-TDR,IceCube-Gen2:2021rkf}. The IceCube-Gen2 radio array will consist of a hybrid array of shallow components with LPDA antennas installed a few meters below the surface and deep components with antennas installed in narrow holes installed down to a depth of \SI{150}{m}. Both components trigger independently. The shallow component requires a high and a low threshold crossing within \SI{5}{ns} with a local coincidence requirement of two out of the four LPDAs fulfilling the trigger condition within \SI{30}{ns}. The deep component is triggered by an interferometric phased array comprising four dipole antennas placed vertically above each other with \SI{1}{m} separation \cite{ARA2019-PA}. Multiple beams covering the relevant zenith angle range are formed by coherently summing the signal from the four antennas to produce synthetic waveforms with increased signal-to-noise ratio. The trigger condition is fulfilled if the integrated power in a sliding \SI{30}{ns} window for any of the synthetic waveforms crosses a certain threshold. The trigger thresholds are adjusted to yield a trigger rate of \SI{100}{Hz} on unavoidable thermal noise fluctuations. In the following, we refer to these two station components as "{LPDA 2of4 100Hz}" and "{PA 4ch 100Hz}". The exact simulation settings and NuRadioMC steering files are available at \cite{Gen2-TDR-NuRadioMC}. In Sec.~\ref{sec:direct_signals}, we will also show results for a phased array at the site of RNO-G. These will use the same frequency band and thresholds but with a depth of \SI{100}{m} and with the commensurate refractive index profile.

We calculate the Askaryan emission using the frequency domain parameterization of in-ice showers \emph{Alvarez2009} \cite{Alvarez-Muniz:2009jsq} as implemented in NuRadioMC \cite{Glaser:2019cws}. We will later correct the effective areas by the correction factor $f$ (see next subsection). 
We describe the index-of-refraction depth profile, i.e., the transition from snow to solid ice in the upper $\mathcal{O}$(\SI{100}{m}), with an exponential model and calculate the signal trajectories through ray tracing \cite{Glaser:2019cws} as done in previous studies, e.g., \cite{IceCube-Gen2-TDR}.

The simulations are performed for reflection layers at \SI{300}{m}, \SI{500}{m}, and \SI{830}{m}. We use the following scheme to directly evaluate all possible reflectivities: In the simulation of the effective area, the reflectivity of these layers was set to unity. Since the reflectivity reduces the signal amplitude linearly, we can infer the rates over a range of reflectivities by simply rescaling the simulated energies.\footnote{For example, a $10^{17}$\,eV shower that triggers a station via a reflection off of a 0\,dB reflector would give an equivalent signal as a $10^{19}$\,eV shower with a -40\,dB reflector.}  As the signal amplitude also depends linearly on the shower energy, we can incorporate any reflectivity later by reducing the effective shower energy accordingly.

\begin{figure}[t]
    \centering
       \includegraphics[width=0.99\columnwidth]{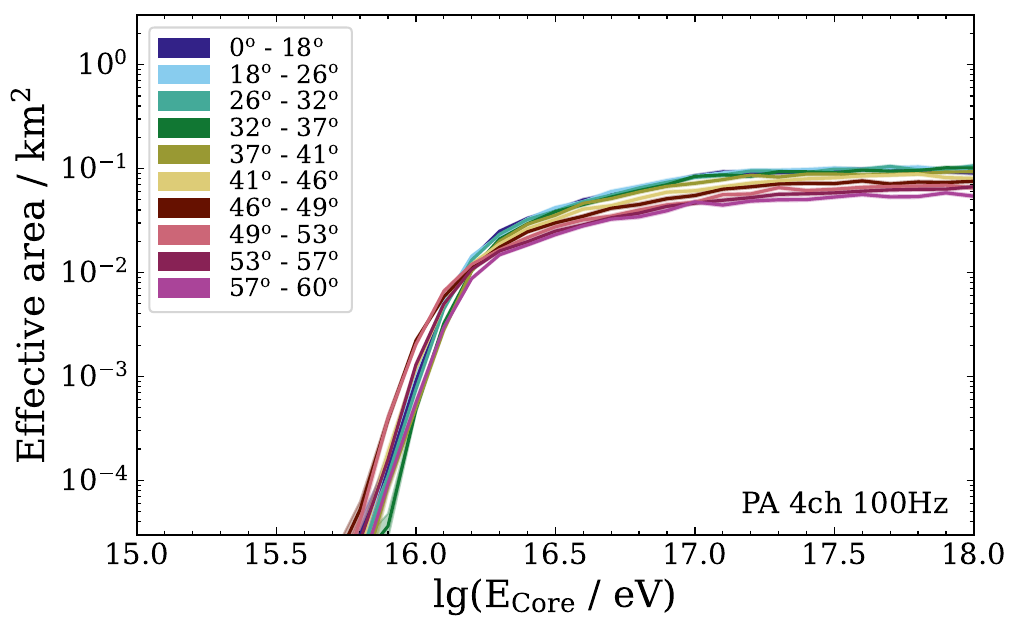}
    \caption{The effective area of detecting air-shower cores impacting the ice through direct signal trajectories with the deep components of an IceCube-Gen2 station. Error bands represent the 68\% statistical uncertainty.}
    \label{fig:effective_area_direct}
\end{figure}

\begin{figure*}[tb]
    \centering
    \includegraphics[width=0.99\columnwidth]{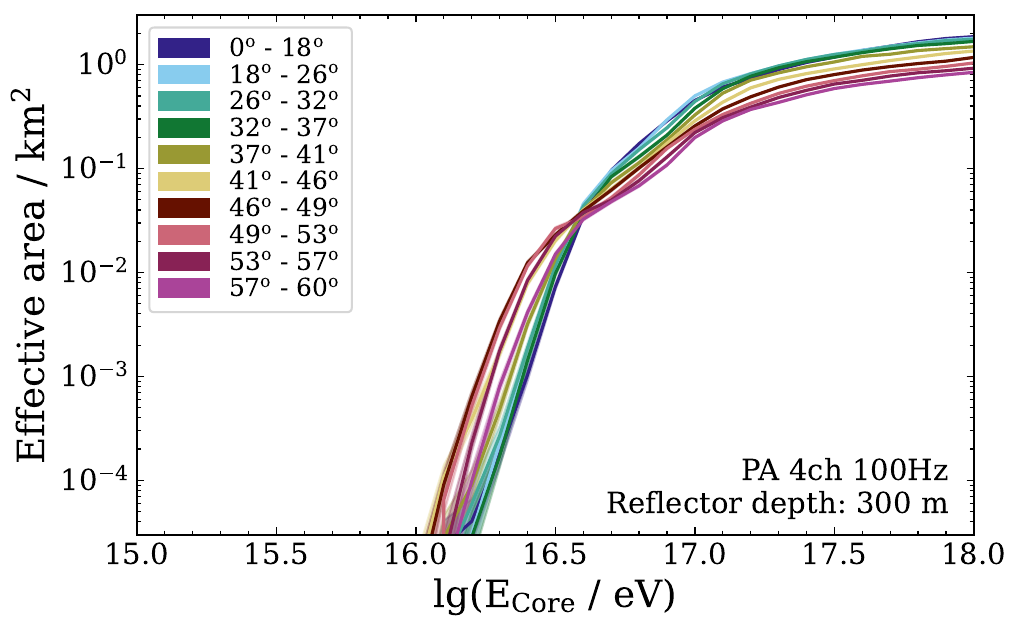}    
    \includegraphics[width=0.99\columnwidth]{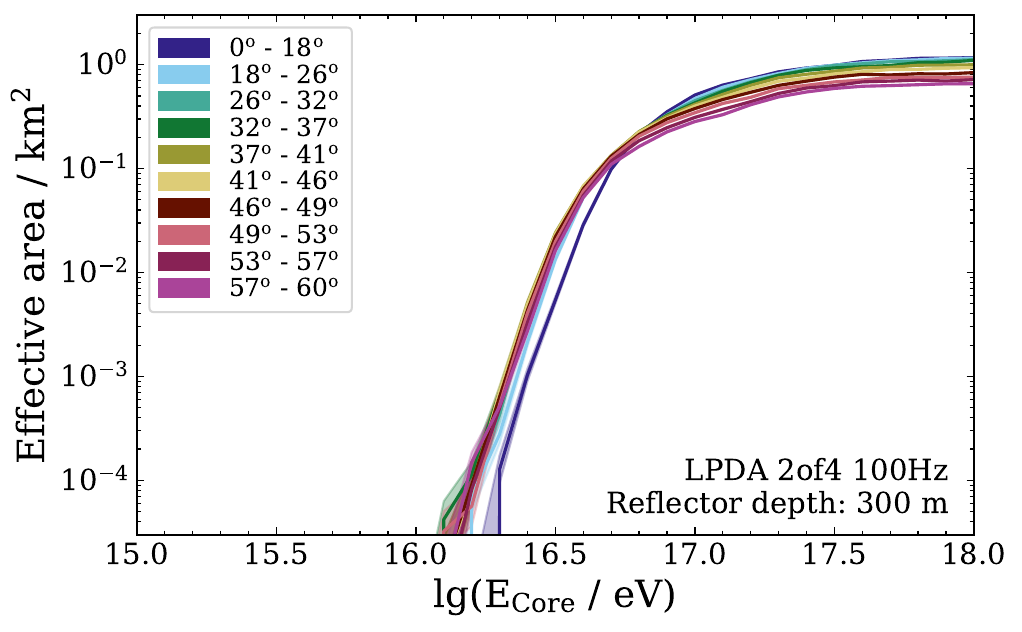}
    \caption{
        The effective area of detecting air-shower cores impacting the ice is shown for the deep (left) and shallow (right) components of an IceCube-Gen2 station. The effective area is shown for events that would be observed in the presence of a 0\,dB reflecting layer at a depth of 300\,m. Error bands represent the 68\% statistical uncertainty.
    }
    \label{fig:effective_area}
\end{figure*}

All showers have an interaction vertex at a depth of \SI{.5}{m} and the simulation is performed for fixed shower energies in steps of 0.1 in $\log_{10}(E)$. The location of the interaction vertex is chosen uniformly on the surface of the ice. The arrival directions are chosen randomly in azimuth and flat in $\cos\theta$, in $\Delta\cos\theta=0.05$ bins. The resulting effective areas are shown in Fig.~\ref{fig:effective_area_direct} and \ref{fig:effective_area} for different zenith angle bands.

\subsection{Air-shower core detection rate}
\begin{figure*}[tb]
    \centering
    \includegraphics[width=0.49\textwidth]{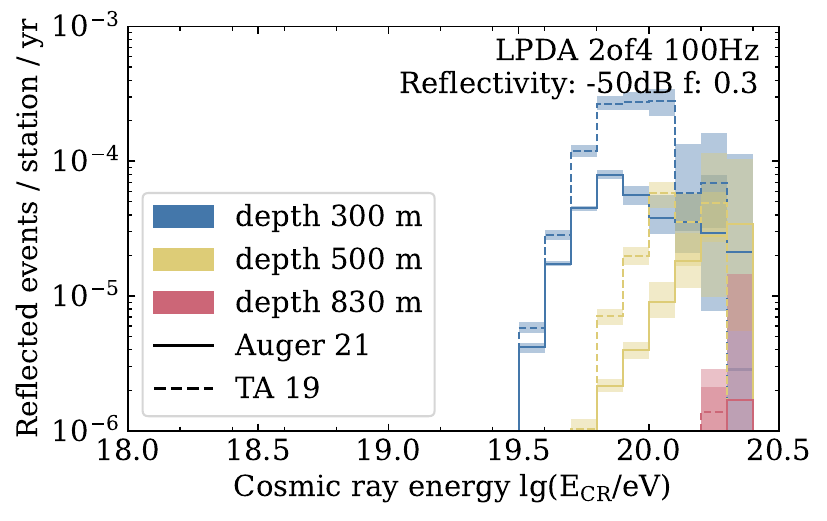}
    \includegraphics[width=0.49\textwidth]{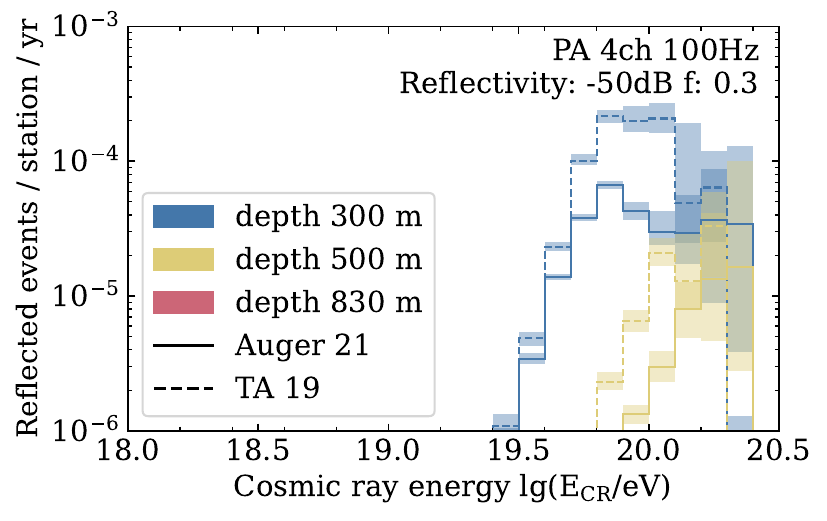}
    \caption{
        The rate of reflected events per year for reflection-layer depths of 300\,m, 500\,m, and 830\,m is shown, assuming the TA and Auger flux for a single IceCube-Gen2 radio station. The rates are made with the assumption of a radio producing fraction $f=0.3$ and a -50\,dB reflector. The rates are shown for the shallow (left) and deep (right) components of an IceCube-Gen2 station. Error bands represent the 68\% interval, taking into account the uncertainties on the cosmic ray flux, the \xmax distributions, and the effective area.
    }
    \label{fig:reflected_events_vs_Ecr}
\end{figure*}

We calculate the detection rate by folding the deposited energy spectrum with the effective area. 
The detection rate of air-shower cores whose signals are reflected up is shown in Fig.~\ref{fig:reflected_events_vs_Ecr} for a reflectivity of \SI{-50}{dB} and correction factor $f$ of 0.3. The detection rate per station is slightly larger for shallow components compared to the deep component (left panel vs. right panel), which is opposite to the case of neutrino detection, where a single deep component has a larger effective area than a single shallow component because it can observe more ice. However, for the special geometry of air shower cores originating at the surface, the emitted signal can always return to the surface after reflection. Hence, the geometrically observable area does not increase by deeper receiver depths. 

We simulated the detection rate for reflection layers at \SI{300}{m}, \SI{500}{m}, and \SI{830}{m}. The shallower the reflection layer, the larger the detection rate, as the signal travels through less ice and is, therefore, less attenuated. We also find that most of the detection rate originates from the highest cosmic-ray energies. Therefore, the differences in the cosmic-ray spectrum at the highest energies between the Auger and TA measurements have a visible impact on the predicted air-shower core detection rates. We also note that the $f$-factor is likely high for these ultra-high-energy air showers (see discussion in Sec.~\ref{sec:C8}).

\begin{figure*}[tb]
    \centering
    \includegraphics[width=1\textwidth]{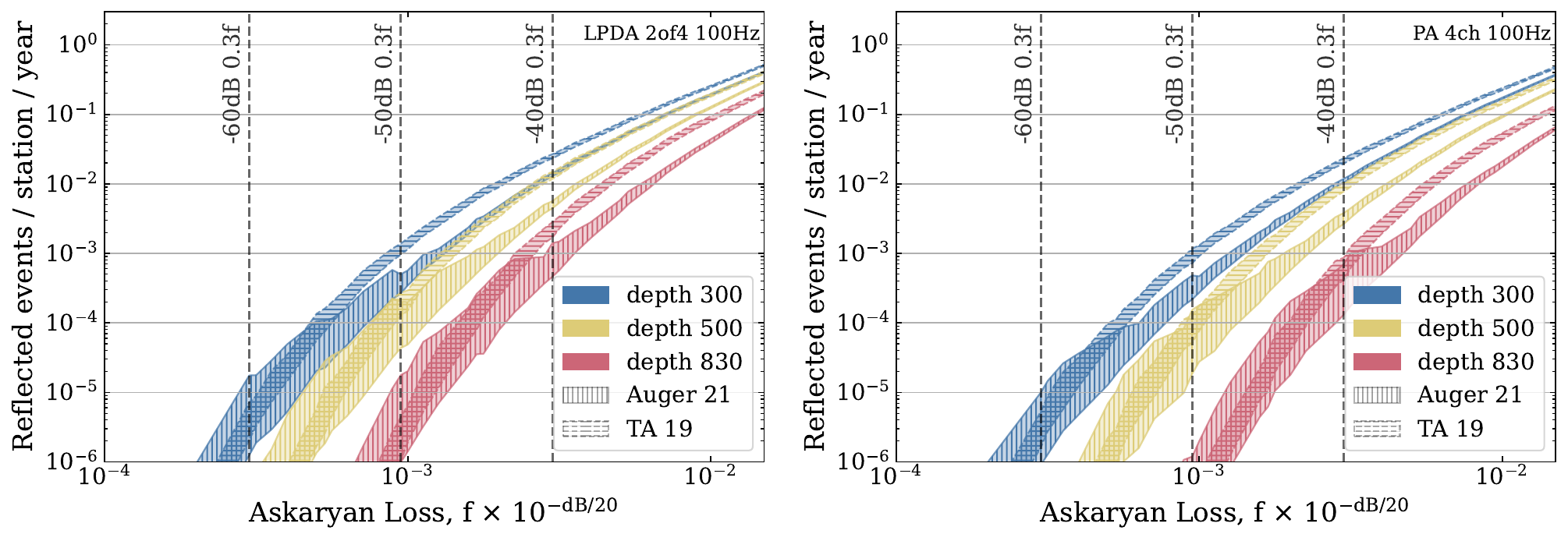}
    \caption{
        The rate of reflected events as a function of the decrease in electric-field strength due to decoherence at the surface ($f$) and losses from the reflection in the deep ice (dB). Several values of the Askaryan loss are shown for $f=0.3$ (vertical dashed gray lines). The bands indicate the 68\% confidence interval, which combines the uncertainties from the measured cosmic-ray flux, \xmax values, and the effective area.
    }
    \label{fig:askaryan_loss_per_station}
\end{figure*}

We integrated over the cosmic-ray energy to calculate the total detection rate of reflected air-shower cores. The resulting event rate per station and year is shown in Fig.~\ref{fig:askaryan_loss_per_station} as a function of the reflectivity and $f$ factor. The plot allows us to read off the background rate for different values of reflectivity and $f$ factor. As also the previous figure showed, the background rate of a shallow detector is slightly higher. In addition, we need to consider that the effective volume of neutrinos (at trigger level) of a single deep detector component is larger than that of a single shallow component by a factor of two to five, depending on the neutrino energy \cite{Barwick:2022vqt}.  Hence, a shallow detector component's background rate normalized to the same neutrino sensitivity is correspondingly larger.

\begin{figure}[tb]
    \centering
    \includegraphics[width=0.99\columnwidth]{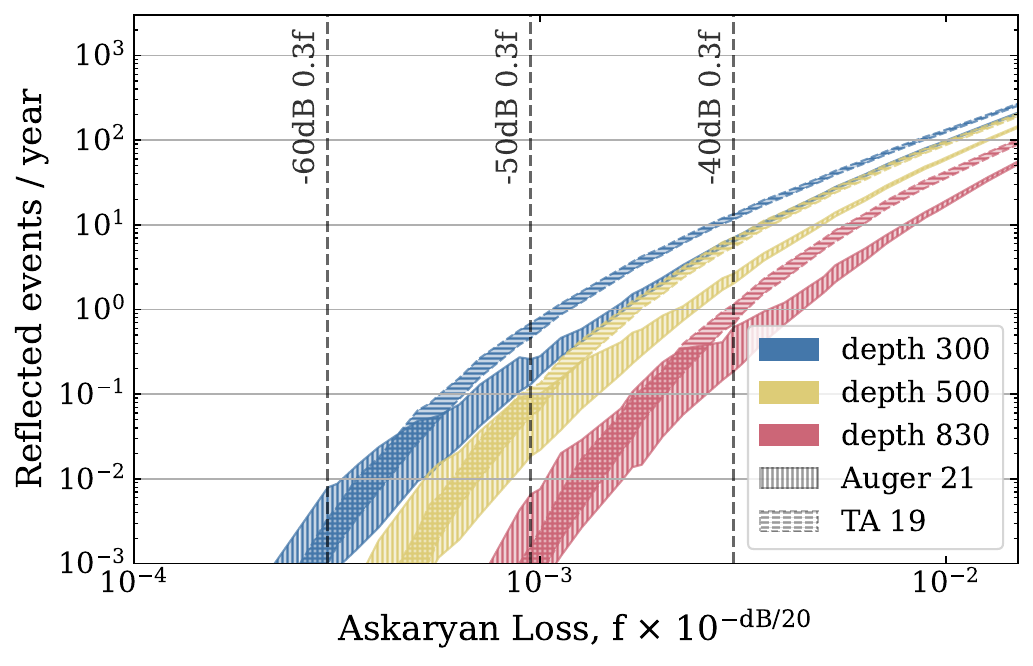}
    \caption{
        The same plot as for~\cref{fig:askaryan_loss_per_station} but for a complete IceCube-Gen2 array with 196 shallow-only stations and 164 hybrid (shallow + deep) stations.
    }
    \label{fig:askaryan_loss_array}
\end{figure}

The IceCube-Gen2 radio array is planned as a hybrid array of 196 shallow-only stations and 164 hybrid stations that contain both a shallow and a deep component. Because each detector station triggers on independent events, and even for a hybrid station, the overlap in triggers between the shallow and deep components is small at low energies, which are relevant for this study (see discussion in Sec.\ref{sec:mitigation} and Fig.~\ref{fig:mitigation_receive}), we can calculate the background rate (per layer) of the entire array by summing the contributions of each station. The resulting rates are shown in Fig.~\ref{fig:askaryan_loss_array}. We find that the depth of the reflection layer strongly affects the background rate, with shallower reflection layers leading to higher background rates. We note that the total background rate is the sum of all layers. 
The relevance of the reflected air-shower core background depends strongly on the reflectivity of the internal ice layers and, to a lesser extent, on the f-factor, which represents our uncertainty of the emission from air-shower cores. For the reflection layer at \SI{500}{m} and a reflectivity of \SI{-50}{dB} and $f$ = 0.3, we expect between 0.02 and 0.1~background events per year across the IceCube-Gen2 array where the uncertainty is coming from the uncertainty of the cosmic-ray energy spectrum and the discrepancy of the Auger and TA measurements (cf. Fig.~\ref{fig:askaryan_loss_array}). As the expected detection rates for neutrinos can be similar depending on the unknown neutrino flux at ultra-high energies (see e.g. \cite{Valera:2022wmu}), the background rate is relevant if it can't be mitigated (see Sec.~\ref{sec:mitigation}). For lower reflectivities of \SI{-60}{dB}, the rate drops to 0.008~events per year, which becomes negligible. For higher reflectivities of \SI{-40}{dB}, which are compatible with the measurement uncertainties of the reflection layers (cf. Sec.~\ref{sec:ice}), the rate approaches several events per year, which is similar or higher than some expectations for neutrino detection rates \cite{Valera:2022wmu}. If a shallower reflection layer at around \SI{300}{m} is present, the background rates increase even further.  We study mitigation strategies in Sec.~\ref{sec:mitigation} and interpret the results in Sec.~\ref{sec:discussion}.

\subsection{Verification through direct signal trajectories}
\label{sec:direct_signals}
The radio emission from air-shower cores is not only a potential background due to the internal reflection layers. It is also a unique calibration source for in-ice radio detectors as we can directly observe the radio emission from air-shower cores with nearby antennas, i.e., without the presence of any reflection layer. This geometry is illustrated as the dashed line in Fig.~\ref{fig:background_diagram}. Deep receivers at $\sim$100~m depth can measure the down-going signal from air-shower cores, i.e., a direct signal trajectory without any reflection at an ice layer. Initially, this can be used to study the radio emission from air-shower cores and validate the prediction from codes such as CORSIKA 8. But once, the air-shower core emission is well understood, it serves as a calibration source. We note that also the in-air generated radio emission from air showers is a useful calibration source, especially for antennas closer to the ice surface as successfully used in several studies \cite{Barwick2017,Arianna:2021lnr,Arianna:2023nvi}.

\begin{figure}[tb]
    \centering
    \includegraphics[width=0.99\columnwidth]{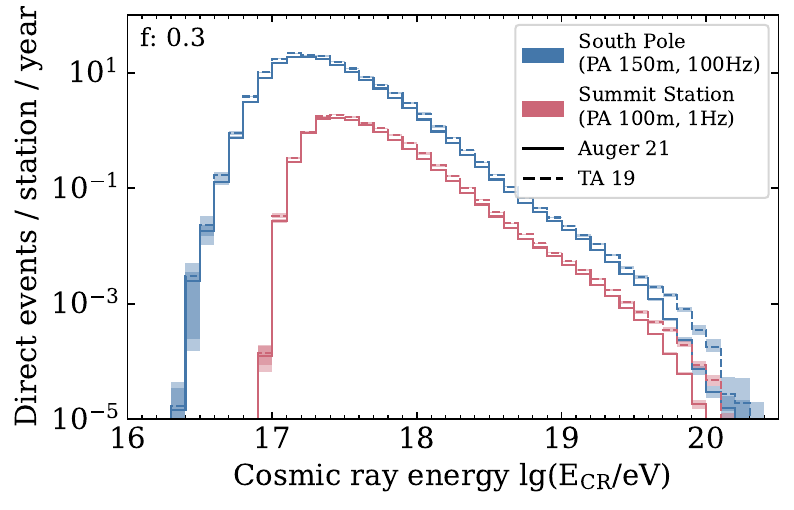}
    \caption{
        The rate of direct-illumination events for a phased-array trigger at a depth of 150\,m at the South Pole (blue) and one at a depth of 100\,m in the Greenland glacier (red). The assumed coherence-loss factor is 0.3. The bands indicate the 68\% confidence interval, which combines the uncertainties from the measured cosmic-ray flux and \xmax values.
    }
    \label{fig:direct_events_vs_Ecr}
\end{figure}

We show the expected detection rates of air-shower cores with direct signal trajectories as a function of cosmic-ray energy in Fig.~\ref{fig:direct_events_vs_Ecr}. We note that the rate prediction only depends on the $f$ factor.
We find that the detection rate at the IceCube-Gen2 detector at the South Pole is significantly larger than for the RNO-G detector at Summit Station in Greenland. This is because the deep component is placed at a depth of \SI{150}{m} at the South Pole and only \SI{100}{m} at Summit Station. In addition, the index-of-refraction profile is different. The transition from snow to deep ice, i.e., the firn, is in the upper \SI{200}{m} at the South Pole, whereas in Greenland, this transition already occurs within the first \SI{100}{m}. The index-of-refraction gradient bends signal trajectories downwards. The RNO-G phased-array antennas are below the firn layer. Hence, the trajectories are maximally curved, which, in turn, minimizes the physical radius between the core impact location and the RNO-G antenna. Therefore, at the South Pole a larger surface area around the deep antennas has direct signal trajectories: For Gen2 at the South Pole, most direct signal trajectories from air-shower cores are observed from a radius of \SI{150}{m} with the distribution extending to over \SI{300}{m}, while for RNO-G at Summit Station Greenland, the distribution peaks at a radius of \SI{80}{m} with the distribution extending to \SI{170}{m}. Because the flux increases linearly with area, the detection rate for Gen2 is larger than for RNO-G. We note that the additional path length to the deeper antenna has negligible impact because the attenuation length is much larger than the path length. 

\begin{figure}[tb]
    \centering
    \includegraphics[width=0.99\columnwidth]{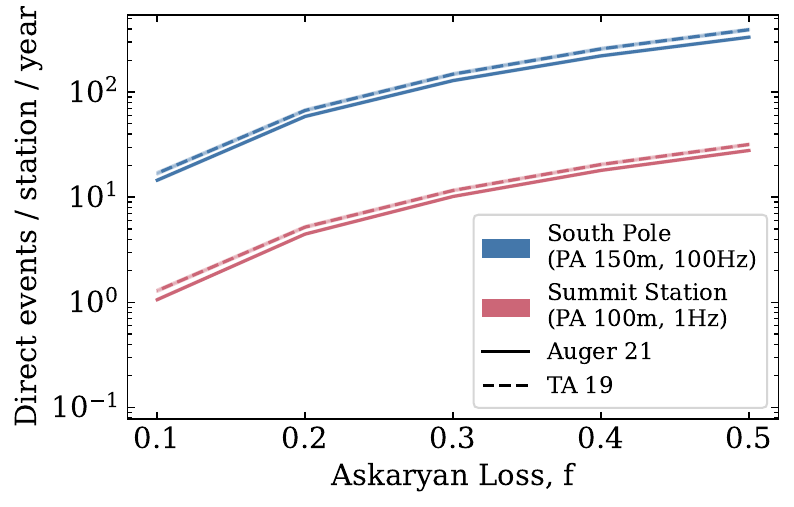}
    \caption{
        The rate of direct-illumination events for a phased-array trigger at a depth of 150\,m at the South Pole (blue) and one at a depth of 100\,m in the Greenland glacier (red). The bands indicate the 68\% confidence interval, which combines the uncertainties from the measured cosmic-ray flux, \xmax values, and the effective area. The assumed coherence-loss factor is 0.3.
    }
    \label{fig:direct_events_askaryan_loss}
\end{figure}

\begin{figure}[tb]
    \centering
    \includegraphics[width=0.99\columnwidth]{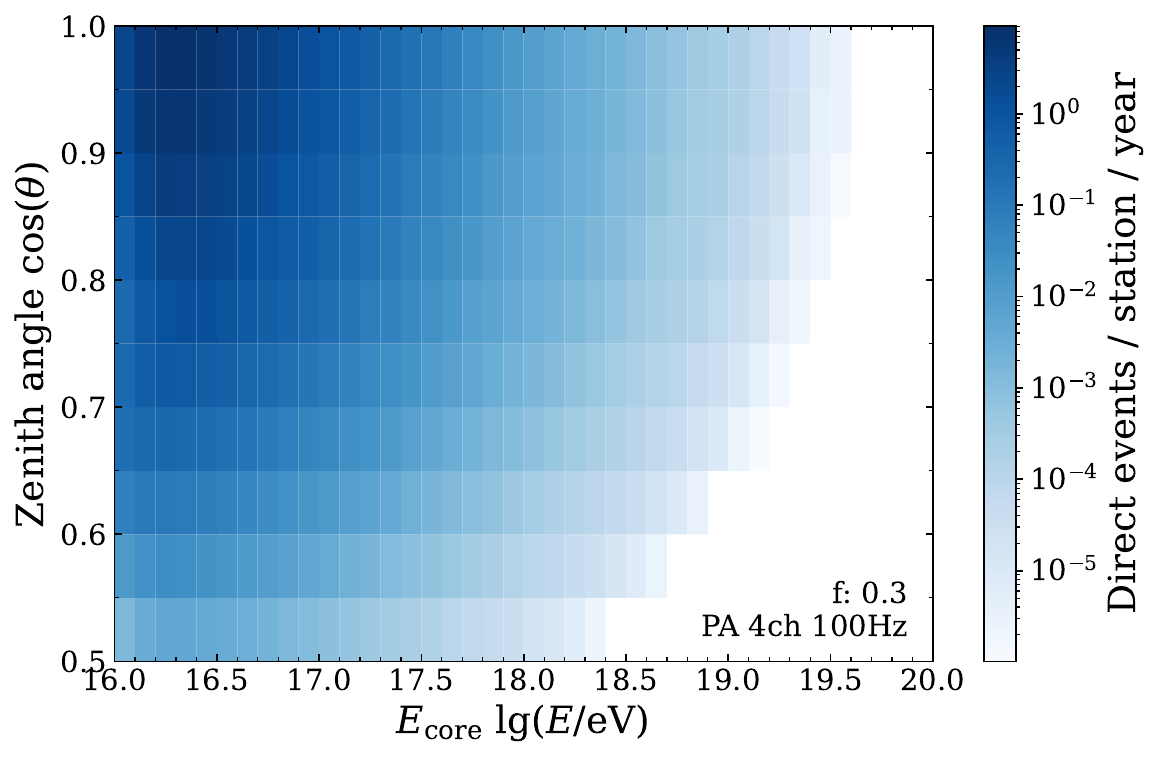}
    \caption{
        The rate of direct-illumination events for the deep component of an IceCube-Gen2 station as a function of \ecore and the zenith angle of the air shower. The assumed coherence-loss factor is 0.3 and the flux is taken to be the one measured by TA.
    }
    \label{fig:direct_events_heatmap}
\end{figure}

The total number of events per year and station as a function of the Askaryan Loss factor $f$ is shown in Fig.~\ref{fig:direct_events_askaryan_loss}. For the reference value of $f=0.3$, we expect 10 detections per year and station at RNO-G, which makes this a clearly detectable signal even with a partly deployed detector. Fig.~\ref{fig:direct_events_vs_Ecr} shows that the rate is dominated by lower energy cosmic rays between \SI{e17}{eV} and \SI{e18}{eV} in contrast to the reflected rates which are dominated by the end of the cosmic-ray spectrum, i.e., energies larger than $10^{19.5}$~eV (cf. Fig.~\ref{fig:reflected_events_vs_Ecr}). Hence, we expect the average $f$-factor to be smaller for the direct-illumination rates, which could lower the rate predictions of Fig.~\ref{fig:direct_events_vs_Ecr}. In the future, we plan to build a better model for the in-ice shower core emission through detailed CORSIKA 8 simulations to decrease this uncertainty.
We also show the rate as a function of the shower-core energy and the air shower's zenith angle in Fig.~\ref{fig:direct_events_heatmap}. Both of these quantities can be reconstructed from measurements and compared to our theory prediction. 

A deep phased array at the South Pole should see ten times more events, i.e., 100 detections per station per year. One station of the existing ARA detector is equipped with a phased array~\cite{ARA2019-PA}, but without having access to details of the ARA signal chain and trigger performance, we can only approximate its performance. The ARA phased array is deployed at a depth of \SI{200}{m} and consists of 7 antennas, which will increase its performance compared to the four-channel phased array at \SI{150}{m} that we simulated in this study. However, the ARA phased-array operates at a much larger bandwidth which decreases its performance, especially to the low-frequency dominated signals of air-shower cores \cite{Glaser2020Bandwidth}. Furthermore, the ARA trigger rate is lower (around \SI{15}{Hz} vs. \SI{100}{Hz}); thus, the thresholds are higher, and the signal chain is more dispersive, which reduces performance compared to our estimates for Gen2. Overall, we expect the ARA phased-array sensitivity to be similar but slightly smaller compared to Gen2.
As the ARA phased array has been running for several years, the experiment could have hundreds of direct detections of air-shower cores in their dataset. So far, no dedicated search for air-shower core signals has been published, but as a side product of the search for UHE neutrinos \cite{ARA:2022rwq}, 46 impulsive events were found originating from the surface region in a half-year data sample \cite{HughesPhD} of which a subset could originate from air-shower cores which is well compatible with our rate estimate, but a more thorough calculation for the ARA detector response and trigger settings are required to solidify the expected detection rate.
Hence, the Askaryan emission from air-shower cores can be probed experimentally using the existing instrumentation or RNO-G or ARA. 

\section{Mitigation Strategies}
\label{sec:mitigation}
If the rate of reflected background events by internal reflectors approaches or exceeds the expected detection rate of EeV neutrinos, which is not excluded within the current range of model uncertainties, then there are a variety of mitigation strategies that can help to distinguish between these two classes of events which we discuss in the following. 

\begin{figure*}[tbp]
    \centering
    \includegraphics[width=0.49\textwidth]{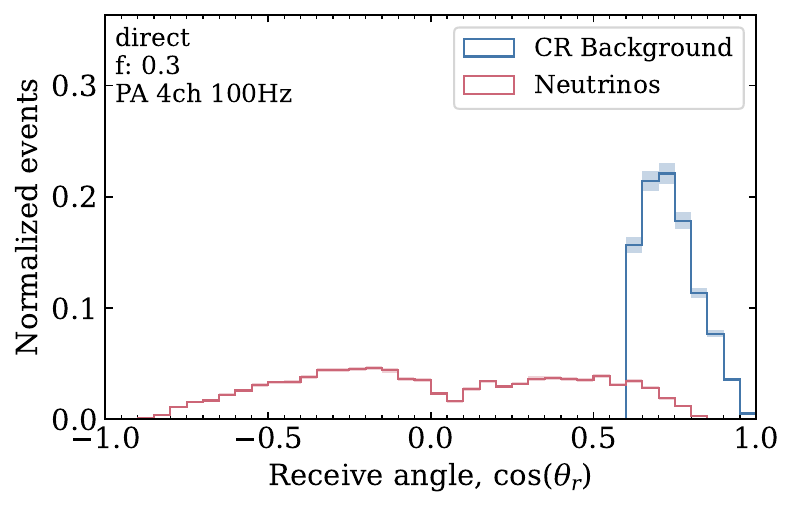}
    \includegraphics[width=0.49\textwidth]{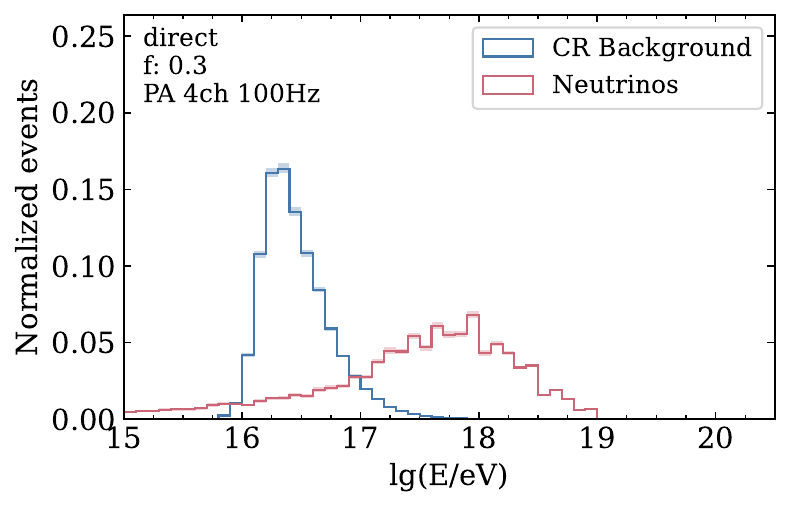}\\
    \includegraphics[width=0.49\textwidth]{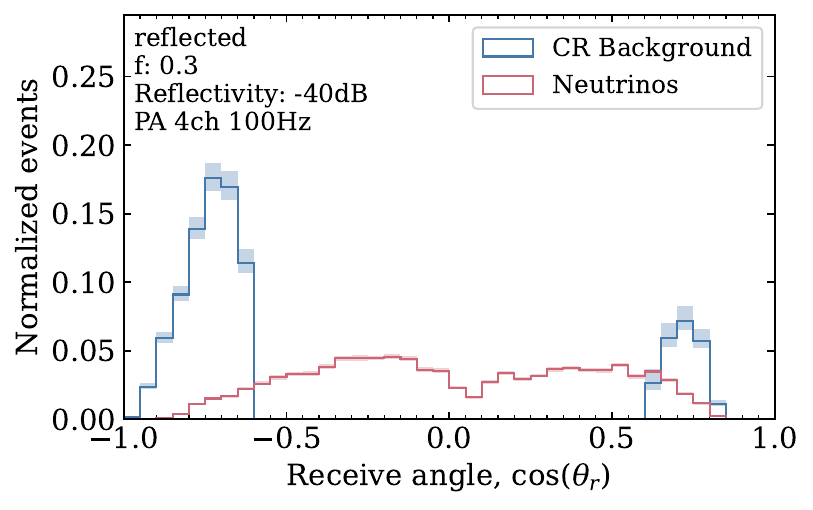}
    \includegraphics[width=0.49\textwidth]{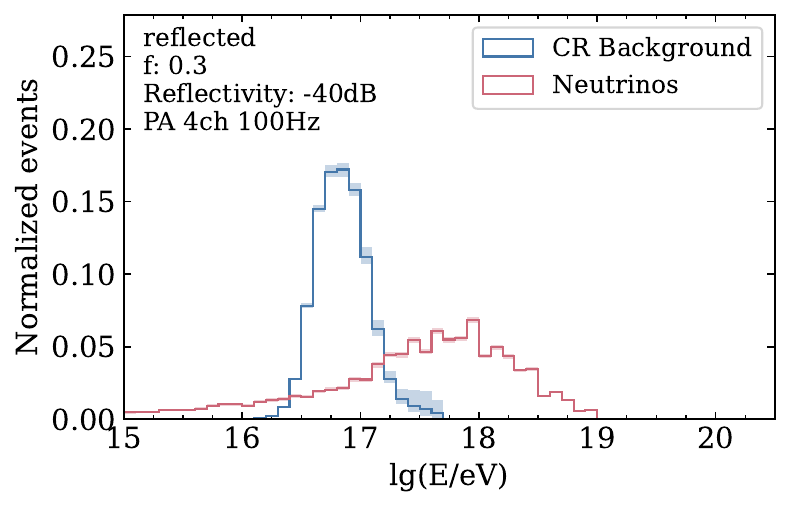}\\
    \includegraphics[width=0.49\textwidth]{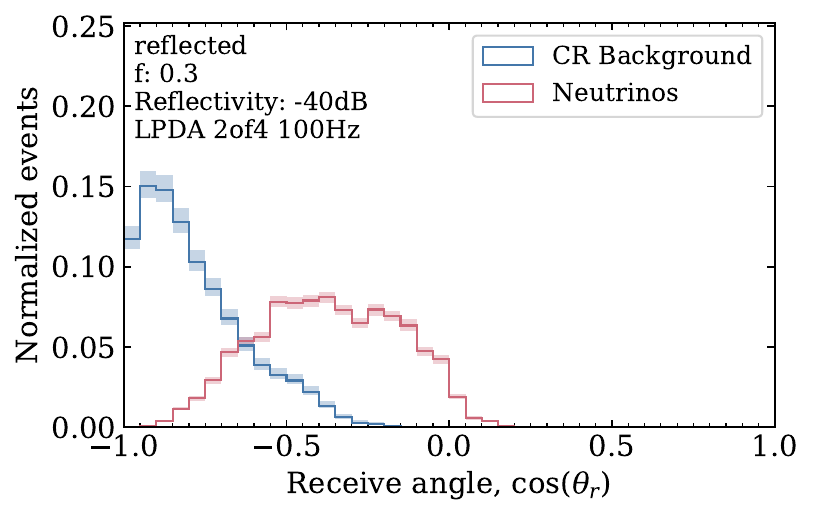}
    \includegraphics[width=0.49\textwidth]{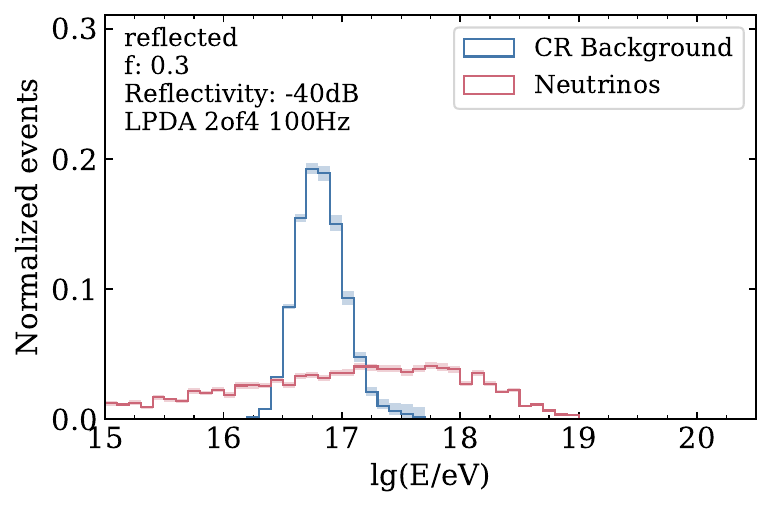}\\
    \caption{
        Signal arrival and shower energy distributions of air-shower core signals and neutrinos. The left panels show the angle from which the shower pulse arrives at the triggering antenna. The right panels show the apparent shower energy. The top panels show the direct trajectories to the deep phased array, the middle panels the reflected trajectories to the deep phased array, and the bottom panels the reflected trajectories to shallow LPDA antennas for a \SI{300}{m} deep reflection layer. The expected distributions for neutrinos are calculated for a typical benchmark flux model of an astrophysical and cosmogenic flux as used in~\cite{Coleman:2024scd} and with an equal mixture of neutrino flavors at Earth.
    }
    \label{fig:mitigation_receive}
\end{figure*} 

\begin{figure*}[tb]
    \centering
    \includegraphics[width=0.49\textwidth]{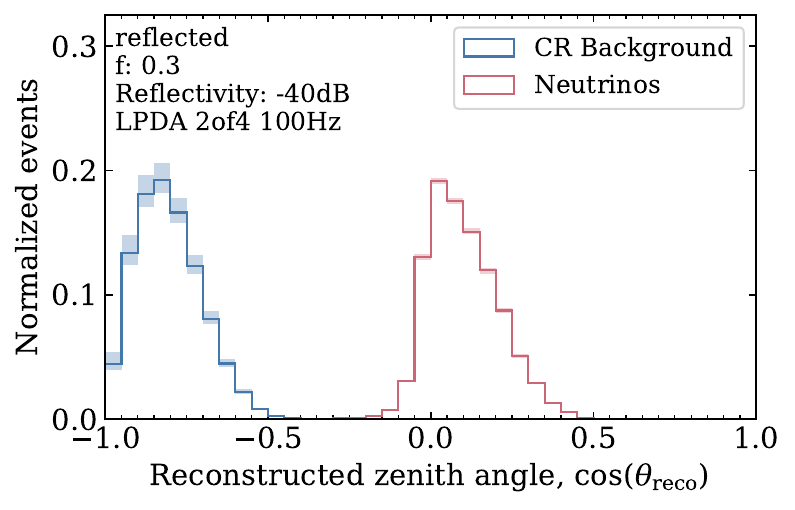}
    \includegraphics[width=0.49\textwidth]{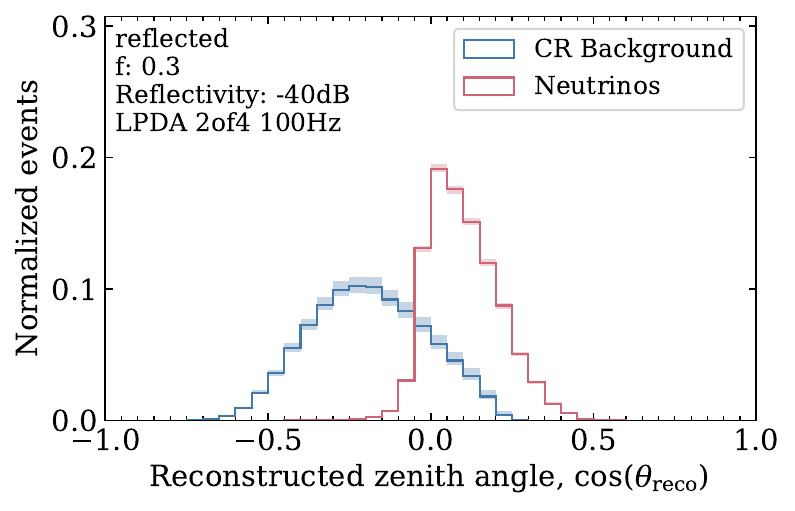} \\
    \includegraphics[width=0.49\textwidth]{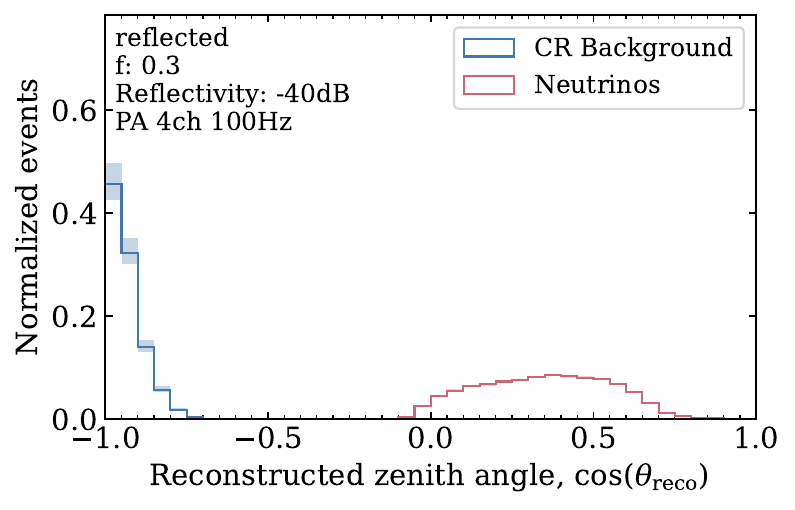}
    \includegraphics[width=0.49\textwidth]{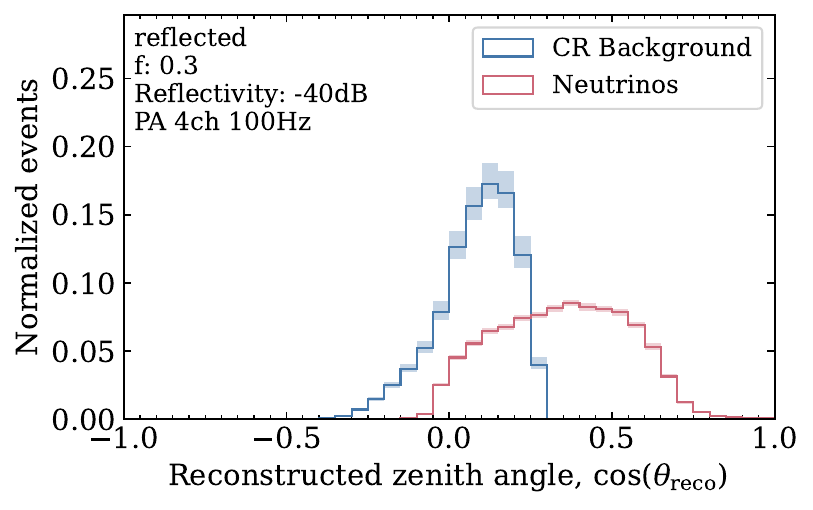}
    \caption{Estimated shower direction, if the direction is reconstructed without knowledge of a reflection layer, is shown in blue, and the expected distribution of neutrino directions is shown in red. Left panels: For reflectors that do not change the phase of the signal. Right panels: For reflectors that change the phase by 180$^\circ$. The upper panels show the distributions for showers triggered by the shallow LPDA antennas, and the bottom panels show the corresponding distributions for the \SI{150}{m} phased array trigger. Only the results for the \SI{300}{m} deep reflector are shown.}
    \label{fig:mitigation_direction}
\end{figure*}  

The first thing that comes to mind is tagging the parent air shower through its in-air-generated radio emission detected by upward-facing LPDA antennas. However, initial studies \cite{ryan_rice_smith_2022_6785120} suggest that the coincident fraction is low because most background events originate from largely vertical air showers where a large fraction of the shower energy is deposited in the ice, which results in little in-air emission distributed over small areas. The reconstruction of the emission point, which is often referred to as interaction vertex position in the case of neutrinos, on the other hand, will be a good discriminator as the vertex of reflected background events will be within 10~m of the snow surface, which would be reconstructed as two times the reflector depth, whereas the vertices of neutrino events will follow a much more diffuse distribution. We note that the reconstruction of the vertex position can serve as a data-driven way of identifying reflection layers for the cases where the detection rate is large. 

Without better modeling of the Askaryan emission from shower cores and a more precise characterization of the reflective properties of the ice layers at the South Pole, the expected effectiveness of mitigation strategies has large uncertainties. We assume that the reflecting layers are smooth and horizontal. However, the layers might not be completely horizontal, which will change the expected signal arrival directions, and non-specular scattering due to roughness may deflect and diffuse the directions, leading to more confusion with neutrino-induced showers. The reflection layer might invert the polarization or not depending on whether the reflectivity originates from an under- or over-density, which will change the reconstructed shower direction. On the other hand, the layers might distort the pulse shape, which will make the waveforms more distinct from the ones we expect from neutrino-induced showers. Similarly, the larger lateral spread of the showers that are induced by air-shower cores compared to the neutrino-induced showers might result in noticeable differences in the pulse shape. However, to use this information, we also need to understand other propagation effects that the South Pole ice has on radio waves, such as birefringence \cite{Heyer:2022ttn}, which potentially distorts neutrino-induced signals that travel for long distances through the ice. For now, we explore the case that the pulse shape cannot be used to identify the background and provide an estimate of how well the background from air-shower cores might be distinguished from neutrinos based on the assumptions of horizontal layers with specular reflection and no dispersion. The development of a vertex reconstruction algorithm is beyond the scope of this paper but we investigate several simpler mitigation strategies to illustrate the the potential effectiveness and experimental issues. 

The simplest quantity that can be reconstructed without any model assumption is the signal arrival direction at the antennas. In Fig.~\ref{fig:mitigation_receive} left, we show the expected distributions in arrival zenith angle for neutrinos and the air-shower core background. We only need to consider the zenith angle because of the cylindrical symmetry of the detector. For the deep antennas, the direct signals originate from above ($\cos\theta$ close to 1). For the reflected signals, most background signals arrive from below ($\cos\theta$ close to -1) but can also arrive from above from signal trajectories that refract downwards or reflect off the surface (often under Total Internal Reflection) at larger distances from the station. In contrast, neutrino signals are more likely to come from more horizontal directions ($\cos\theta$ close to 0). For the shallow antennas, the background is expected to arrive only from below ($\cos\theta$ close to -1), and again, the neutrino signals generally arrive from more horizontal directions. In both cases, a cut in the receive angle can reduce large fractions of the background while maintaining a large fraction of the neutrino signals. The cut is more efficient for deep than for shallow detector components. 

In addition, we calculated the expected shower energy distribution, which might provide an additional handle to reject the background. For the reflected signals from air-shower cores, we calculate the apparent shower energy \eapparent, i.e., the true shower energy reduced by the f-factor and the reflectivity, because this is the shower energy a reconstruction would estimate as we have no prior knowledge of whether the signal was reflected or not. Hence, the distribution \eapparent depends on the assumption of reflectivity and f-factor. Here, we show the distribution for a reflectivity of \SI{-40}{dB} and $f=0.3$. Smaller/larger reflectivities will shift the distribution linearly to lower/higher energies. The separability of neutrinos and air-shower core background using the reconstructed shower energy depends on the reflectivity (and f-factor) but would generally limit the neutrino detection efficiency significantly.

More discrimination power can be achieved by also taking the signal polarization into account to determine the shower direction. The shower (or neutrino) direction is given by \cite{Barwick:2022vqt}
\begin{equation}
    \vec{\hat{v}}_\nu = \sin \theta \vec{\hat{p}} + \cos \theta \, \vec{\hat{l}} \, ,
\end{equation}
where the $\hat{}$ symbol indicates that the vectors have unit length, $\theta$ is the viewing angle, $\vec{\hat{p}}$ is the signal polarization, and $\vec{l}$ is the launch vector which corresponds to the incoming signal direction after correcting for the bending in the firn. We assume perfect knowledge of the signal arrival direction, polarization, and viewing angle to explore how well the signal and background distributions separate. In reality, the distributions will be smoothed out further by reconstruction uncertainties. The result is shown in Fig.~\ref{fig:mitigation_direction}. The separability crucially depends on whether the polarization is inverted upon reflection. For reflection layers that do not invert the polarization, the shower direction distributions separate perfectly (left panels). As expected, the reflected air-shower core signals are reconstructed as arriving from below, i.e., the true air-shower direction is mirrored horizontally. However, if the reflection layer inverts the polarization, the air-shower core signals are reconstructed as arriving from horizontal directions and the two distributions do not separate well anymore (right panels).

Another potential option to reject the air-shower-core background is by \emph{horizontal propagation}. It was found that a small fraction of radio-signal power can propagate horizontally through the ice and does not get bent downwards by the index-of-refraction gradient of the upper firn layer \cite{Barwick2018,Deaconu2018}. A small part of the radio emission generated by the air-shower cores can couple into these horizontal propagation modes, travel large distances, and might be observed by the shallow LPDA antennas of a radio detector station. The recent development of the Eisvogel code \cite{Windischhofer:2023ahw} and its ongoing integration into CORSIKA8 \cite{Alameddine:2024cyd} will allow theoretical calculation of this emission mode without any simplifications of the propagation of radio waves in ice, such as the typical use of ray tracing. Instead, Maxwell's equations are evolved in time using a finite-difference time-domain (FDTD) code such as MEEP \cite{OSKOOI2010687}.
We would like to point out that this discussion has probably not exhausted possible mitigation strategies. 

\section{Discussion and Outlook}
\label{sec:discussion}

Ultra-high-energy neutrinos can be detected through in-ice radio detector stations installed in large ice sheets that search for the Askaryan emission from neutrino-induced in-ice showers. Typical sites are the South Pole, the place of the envisioned IceCube-Gen2 observatory, and central Greenland, home to the RNO-G detector, which are at high elevations of almost 3000~m above sea level. At these high altitudes, cosmic-ray-induced air showers will deposit a large fraction of their energy into the ice, initiating in-ice showers that generate radio emission through the same process as neutrino-induced in-ice showers. Here, we show, through novel CORSIKA 8 simulations, that the air-shower cores generate coherent in-ice radio emission similar to a neutrino-induced in-ice shower of the same energy. For nearby air-shower cores, their Askaryan emission can be picked up by the deep antennas of an in-ice radio detector station, especially by antennas of the interferometrical phased array installed at approx.~150~m depth (for Gen2). We calculated the detection rate for the first time and estimated that IceCube-Gen2 should detect around a hundred events per station per year. Similar detection rates are expected for the ARA detector station at the South Pole, which is equipped with a phased-array trigger. For RNO-G, we estimate around ten detections per detector station and year due to the different ice, shallower antenna positions, and difference in the trigger. Hence, this is a so far under-appreciated calibration source that will allow the verification of most analysis steps needed for neutrino detection: triggering, signal identification, and reconstruction of the shower direction and energy. Furthermore, the radio emission from air-shower cores is already testable experimentally with data from the ARA and RNO-G detectors.  

In addition, the presence of reflection layers at the South Pole makes the Askaryan emission from air-shower cores a potentially relevant background to UHE neutrino detection. Because the Askaryan signal is reflected back up, it can be confused with Askaryan emission from neutrinos. Our background rate estimates for the IceCube-Gen2 array have large uncertainties and range from negligible detection rates (much less than one detection in ten years) to several detections per year, which would be relevant if not mitigated. The background rate normalized to the neutrino sensitivity is higher for the shallow detector component. 
The main uncertainties in our prediction and characterization of the background are the modeling of the radio emission from air-shower cores and the properties of the reflection layers at the South Pole. The results presented here provide strong motivation to study these two aspects in more detail. The uncertainty of the shower-core radio emission can be reduced on a comparably short timescale with existing resources and instrumentation. Reducing the uncertainty on reflection layers would require a new measurement campaign at the South Pole. We will discuss this in the following. 

The ongoing development of the CORSIKA~8 code will allow the simulation of particle showers in arbitrary media and the resulting radio emission by tracking each individual shower particle~\cite{Engel:2018akg,CORSIKA:2023jyz,Alameddine:2024cyd}. As radio emission is calculated from first principles, the theoretical uncertainties are small, as already demonstrated for radio emission from air showers (e.g. \cite{GlaserErad2016,Gottowik:2017wio,SlacT510}). In particular, it allows us to simulate the complex geometries of air showers hitting the ice and the propagation of the generated radio emission through inhomogeneous media such as the ice. In this article, we have already shown a result from CORSIKA~8 for a simplified geometry with homogeneous ice, adequate for comparison to the radio-emission models of neutrinos (see \cite{Alvarez_box,Alvarez2009}). Two of the authors (AC, CG) are involved in the CORSIKA~8 development and will integrate the missing ingredient of tracking radio emission through inhomogeneous media. Then, the relevant geometries can be simulated with high accuracy. 
Direct integration of CORSIKA 8 into NuRadioMC to calculate the effective area to shower cores is unfeasible due to the long runtimes of CORSIKA 8 of several core days per shower, but we can use CORSIKA 8 to parameterize the emission as was done previously for neutrino-induced in-ice showers (see e.g. \cite{Alvarez_box,Alvarez2009,Alvarez-Muniz:2011wcg}). Neutrino-induced Askaryan emission is characterized well by only two variables, the shower energy and the viewing angle. We expect that the Askaryan emission from shower cores can be parameterized by adding the cosmic-ray energy, or rather the ratio between the in-ice shower and cosmic-ray energy (which implicitly takes the zenith angle dependence into account), as an additional variable. Such a fast parameterization will remove the \emph{Askaryan loss $f$-factor} and its associated uncertainty from our analysis. 

With data from the RNO-G observatory, it will be possible to measure the shower core emission experimentally via direct signal trajectories to the 100~m deep antennas, i.e., without any reflection. We predict that RNO-G should measure around 10 air-shower cores per year and station. With eight stations operational, RNG-G should see hundreds of these events.
We note that the radio emission generated in the air by air showers will also be visible in the in-ice antennas, and a selection of air shower core signals would need to discriminate this background by using, e.g., signal polarization as a discriminator. 
It will be interesting to study how often the upward-facing LPDA antennas will also observe the in-air generated radio emission, which will provide an independent measurement of the air-shower direction as well as an estimate of the deposited shower energy in the atmosphere. Such a combined measurement will allow to verify the calculation of air-shower core emission experimentally. 
We note that once the air-shower core emission is understood, a combined measurement of the in-air and in-ice radio emission can be used to estimate the cosmic-ray energy. The sum of the deposited energy in air (estimated by the in-air radio emission \cite{Welling2019} and in ice (estimated by the in-ice radio emission \cite{Aguilar:2021uzt, Glaser:2022lky}) corresponds to the complete electromagnetic shower energy, which is a good proxy for the cosmic-ray energy. Furthermore, such a combined measurement might enable interesting air-shower physics and an estimation of the cosmic-ray composition, as the ratio of in-air and in-ice shower energy is sensitive to the longitudinal shower profile. 

To reduce the uncertainty on the properties of the reflection layers, a new measurement campaign at the South Pole is needed. 
In general, there are two approaches to extracting the properties of internal layers such as absolute reflectivity, signal dispersion/distortion, and phase shifts. The first requires precise knowledge of all elements of full-circuit RF gain, including transmitters and receivers, allowing extraction of reflection coefficients based on the observed amplitude of reflected signals as recently performed at the RNO-G site~\cite{RNO-G:2023Ice}. The second uses the observed amplitude of an internal-layer reflection relative to an observed reflector of known strength (e.g., the signal from an in-ice transmitter reflected at the ice-air boundary or at the bedrock) to quantify the former reflection coefficient. Previous estimates at both the South Pole and also Summit Station have employed both techniques, albeit using transmitters and receivers deployed on the surface, which increased systematic uncertainties due to the modeling of the coupling into the ice. A future dedicated campaign at South Pole could achieve high ($<$3\,dB) precision on the absolute reflectivity of the shallowest bright reflecting layer ($z\sim -$300\,m), provided a dedicated, calibrated transmitter can be lowered into the SPICE borehole, and optimally-positioned receivers can be installed below the ice surface. 

In summary, impacting air-shower cores at high-elevation sites generate significant in-ice radio emission through the Askaryan effect. With tens of detections per year per detector station through direct signal trajectories that can be discriminated by their signal arrival direction from neutrino-induces Askarayn emission, air-shower core signals are a great calibration source for in-ice radio neutrino detectors. If strong enough internal reflection layers exist, these signals can become a relevant background. The calculated background rates, given the current uncertainties, vary from negligible to highly significant if not mitigated. Initial studies of mitigation strategies are promising, but better characterization of the reflection layers is required to make firm statements.

\section*{Acknowledgements}
We thank the CORSIKA~8 collaboration for their support. We thank Philipp Windischhofer for helpful discussions about Fresnel reflection coefficients. CG and AC are supported by the Swedish Research Council {\sc (Vetenskapsrådet)} under project no.~2021-05449. CG is supported by the European Union (ERC, NuRadioOpt, 101116890). This work used resources provided by the National Academic Infrastructure for Supercomputing in Sweden (NAISS) and the Swedish National Infrastructure for Computing (SNIC) at UPPMAX partially funded by the Swedish Research Council through grant agreements no. 2022-06725 and no. 2018-05973. DB is supported by the National Science Foundation's generous IceCube EPSCoR Initiative  grant \#2019597.

\bibliographystyle{JHEP}
\bibliography{bib}

\end{document}